\documentclass[preprint,12pt]{elsarticle}

\usepackage{amssymb}
\usepackage{bm}
\usepackage{multirow}
\usepackage{multicol}
\usepackage{adjustbox}
\usepackage{amsmath}
\usepackage{cases}
\usepackage{svg}
\usepackage{lineno}
\usepackage{caption}
\usepackage{subcaption}

\begin{document}
\begin{frontmatter}

\title{CFD-based Design Optimization of Ducted Hydrokinetic Turbines}

\author[inst1]{Jeongbin Park}
\author[inst1]{Bradford G. Knight}
\author[inst2]{Yingqian Liao}
\author[inst2]{Marco Mangano}
\author[inst2]{Bernardo Pacini}
\author[inst1]{Kevin J. Maki}
\author[inst2]{Joaquim R. R. A. Martins}
\author[inst1]{Jing Sun}
\author[inst1]{Yulin Pan\corref{cor1}}
\ead{yulinpan@umich.edu}
\cortext[cor1]{Corresponding author}

\affiliation[inst1]{organization={Naval Architecture and Marine Engineering, University of Michigan},
            city={Ann Arbor},
            postcode={MI 48109}, 
            state={MI},
            country={USA}}
\affiliation[inst2]{organization={Aerospace Engineering, University of Michigan},
            city={Ann Arbor},
            postcode={48109}, 
            state={MI},
            country={USA}}

\begin{abstract}
Hydrokinetic turbines extract kinetic energy from moving water to generate renewable electricity, thus contributing to sustainable energy production and reducing reliance on fossil fuels. It has been hypothesized that a duct can accelerate and condition the fluid flow passing the turbine blades, improving the overall energy extraction efficiency. However, no substantial evidence has been provided so far for hydrokinetic turbines.
To investigate this problem, we perform a CFD-based optimization study with a blade-resolved Reynolds-averaged Navier--Stokes (RANS) solver to explore the design of a ducted hydrokinetic turbine that maximizes the efficiency of energy extraction. To handle the high-dimensional design space of the blade and duct geometry, we use a gradient-based optimization approach where the gradients are computed using the adjoint method. The final design is re-evaluated through higher-fidelity unsteady RANS (URANS) simulations. Our optimized ducted turbine achieves an efficiency of about 54\% over a range of operating conditions, higher than the typical 46\% efficiency of unducted turbines such as the well-known Bahaj model~\cite{bahaj2007power}.   
\end{abstract}





\end{frontmatter}

\section{Introduction}
\label{sec:intro}

The increasing demand for renewable energy has motivated extensive research on hydrokinetic energy conversion systems that extract energy from natural riverine and oceanic flows. Various types of conversion systems have been investigated for decades, including horizontal- and vertical-axis turbines and oscillating hydrofoils~\cite{khan2009hydrokinetic,lago2010advances,ibrahim2021hydrokinetic}.
Horizontal-axis turbines have been studied the most because of the relatively mature technology~\cite{corsatea2013overview,magagna2015ocean, winslow2017urban,yahyaoui2018modeling,eriksson2008evaluation}. 

A popular benchmark for horizontal-axis hydrokinetic turbines is the Bahaj model, which has been experimentally tested in a cavitation tunnel and a towing tank~\cite{bahaj2007power}. The unducted Bahaj model generates power with an efficiency of about 46\% (the ratio of generated power to the inflow power) at the optimal operating condition.
This is the typical efficiency level of well-designed hydrokinetic turbines~\cite{NUNES2020}. To evaluate this efficiency, we can compare it to the well-known Betz’s limit of 59.3\%~\cite{betz1920maximum}, which is derived based on the one-dimensional (1D) momentum theory in an unbounded flow domain.
There is no general consensus on whether the Betz’s limit should be considered as a hard upper bound on the efficiency of practical energy conversion systems in unbounded flow due to simplifications in the theory. However, it seems clear that further improvement can be sought regarding the current 46\% efficiency of horizontal-axis hydrokinetic turbines.

One idea to improve the efficiency is to use a duct (also known as a shroud or diffuser) to accelerate the fluid flow passing the turbine blades, thus improving the efficiency of the device.
Some researchers have incorporated the duct effect in the 1D momentum theory (or its extended version), with some of which predicting an efficiency well above the Betz’s limit~\cite{lilley1956preliminary, igra1981research, vries1979fluid, lawn2003optimization, van2007science, jamieson2009beating, werle2008ducted, Knight2018, werle2020enhanced, bontempo2020diffuser}. In spite of the insight into the duct effects, the physics may be oversimplified (sometimes misrepresented) meaning the efficiency predicted by these models may not be achievable in practice. The complex turbine-duct interaction involves flow features, such as flow separation, that cannot be captured by the analytical models. These phenomena can significantly affect the mass flow through the duct and the system efficiency~\cite{Knight2018}.

\begin{table}[h!]
\centering
\begin{adjustbox}{width=\columnwidth,center}
\begin{tabular}{lllllrc}
    \multirow{2}{*}{\textbf{\large Researchers}} & \multirow{2}{*}{\textbf{\large Type}} & \multirow{2}{*}{\textbf{\large Shape}} & \multirow{2}{*}{\textbf{\large Method}} & \multirow{2}{*}{\textbf{\large $C_{P,A_b}$}} & \multirow{2}{*}{\textbf{\large $C_{P,A_\text{max}}$}} & \multirow{2}{*}{\textbf{\large $Re$}}\\ &&&&&& \\
    \hline

    \multirow{2}{*}{\citet{aranake2017aerodynamic}} & \multirow{2}{*}{Wind} & \multirow{2}{*}{Foil} & \multirow{2}{*}{Numerical (RANS + BET)} & \multirow{2}{*}{\large 1.57} & \multirow{2}{*}{\large 0.85} & \multirow{2}{*}{$\sim5\times 10^{6}$}\\ &&&&&& \\ 
    
    \multirow{2}{*}{\citet{venters2018ducted}} & \multirow{2}{*}{Wind} & \multirow{2}{*}{Foil} & \multirow{2}{*}{Numerical (RANS + actuator disk)} & \multirow{2}{*}{--} & \multirow{2}{*}{\large 0.62} & \multirow{2}{*}{$\sim1.2\times 10^{6}$}\\ &&&&&& \\ 

    \multirow{2}{*}{\citet{roshan2015rans}} & \multirow{2}{*}{Wind} & \multirow{2}{*}{Stepped} & \multirow{2}{*}{Experimental and numerical (Blade-resolved RANS)} & \multirow{2}{*}{\large 0.86} & \multirow{2}{*}{\large 0.44} & \multirow{2}{*}{$\sim3\times 10^{6}$}\\ &&&&&& \\ 
    
    \multirow{2}{*}{\citet{ohya2010shrouded}} & \multirow{2}{*}{Wind} & \multirow{2}{*}{Flanged} & \multirow{2}{*}{Experiment} & \multirow{2}{*}{--} & \multirow{2}{*}{\large 0.48-0.54} & \multirow{2}{*}{$\sim5\times 10^{5}$}\\ &&&&&& \\ 

    \multirow{2}{*}{\citet{oka2015aerodynamic}} & \multirow{2}{*}{Wind} & \multirow{2}{*}{Flanged} & \multirow{2}{*}{Experiment} & \multirow{2}{*}{--} & \multirow{2}{*}{\large 0.62} & \multirow{2}{*}{$\sim5\times 10^{5}$}\\ &&&&&& \\ 

    \multirow{2}{*}{\citet{oka2015aerodynamic}} & \multirow{2}{*}{Wind} & \multirow{2}{*}{Flanged} & \multirow{2}{*}{Numerical (RANS + BEMT)} & \multirow{2}{*}{\large 1.00} & \multirow{2}{*}{\large 0.62} & \multirow{2}{*}{$\sim8\times 10^{5}$}\\ &&&&&& \\ 

    \multirow{2}{*}{\citet{kosasih2012experimental}} & \multirow{2}{*}{Wind} & \multirow{2}{*}{Flanged} & \multirow{2}{*}{Experiment} & \multirow{2}{*}{--} & \multirow{2}{*}{\large 0.41-0.52} & \multirow{2}{*}{$\sim1\times 10^{5}$}\\ &&&&&& \\ 

    \multirow{2}{*}{\citet{khamlaj2018analysis}} & \multirow{2}{*}{Wind} & \multirow{2}{*}{Flanged} & \multirow{2}{*}{Numerical (RANS + BEMT)} & \multirow{2}{*}{--} & \multirow{2}{*}{\large 0.50} & \multirow{2}{*}{$\sim6.3\times 10^{5}$}\\ &&&&&& \\ 

    \multirow{2}{*}{\citet{heikal2018actual}} & \multirow{2}{*}{Wind} & \multirow{2}{*}{Flanged} & \multirow{2}{*}{Numerical (Blade-resolved RANS)} & \multirow{2}{*}{--} & \multirow{2}{*}{\large 0.57} & \multirow{2}{*}{$\sim1\times 10^{6}$}\\ &&&&&& \\ 
    
    \multirow{2}{*}{\citet{gaden2010numerical}} & \multirow{2}{*}{Hydro} & \multirow{2}{*}{Thin} & \multirow{2}{*}{Numerical (RANS + BEMT)} & \multirow{2}{*}{\large 0.87} & \multirow{2}{*}{\large 0.54} & \multirow{2}{*}{$\sim9\times 10^{6}$}\\ &&&&&& \\ 
    
    \multirow{2}{*}{\citet{coiro2016diffuser}} & \multirow{2}{*}{Hydro} & \multirow{2}{*}{Foil} & \multirow{2}{*}{Experiment and numerical (RANS + actuator disk, blade-resolved)} & \multirow{2}{*}{\large 0.78} & \multirow{2}{*}{\large 0.44} & \multirow{2}{*}{$\sim3\times 10^{6}$}\\ &&&&&& \\ 
    
    \multirow{2}{*}{\citet{tampier2017numerical}} & \multirow{2}{*}{Hydro} & \multirow{2}{*}{Foil} & \multirow{2}{*}{Numerical (Blade-resolved URANS)} & \multirow{2}{*}{\large 0.61} & \multirow{2}{*}{\large 0.19} & \multirow{2}{*}{$\sim7\times 10^{6}$}\\ &&&&&& \\ 
    
    \multirow{2}{*}{\citet{Knight2018}} & \multirow{2}{*}{Hydro} & \multirow{2}{*}{Thin} & \multirow{2}{*}{Numerical (Blade-resolved RANS and URANS)} & \multirow{2}{*}{--} & \multirow{2}{*}{\large 0.45} & \multirow{2}{*}{$\sim2\times 10^{6}$}\\ &&&&&& \\ 
    
    \multirow{2}{*}{\citet{song2019numerical}} & \multirow{2}{*}{Hydro} & \multirow{2}{*}{Foil} & \multirow{2}{*}{Numerical (Blade-resolved RANS)} & \multirow{2}{*}{\large 0.67} & \multirow{2}{*}{\large 0.38} & \multirow{2}{*}{$\sim3\times 10^{6}$}\\ &&&&&& \\ 
    
    \hline \hline
\end{tabular}
\end{adjustbox}
\caption{Previous publications on ducted wind and hydrokinetic turbines. For research evaluating the performance with blade swept area as the reference area, we provide the values of $C_{P,A_b}$ and the converted values $C_{P,A_\text{max}}$ (which we calculate when necessary). The Reynolds numbers are based on the maximum diameter of a whole device as defined in Eq.~\eqref{eq_TSR}. BEMT is the blade element momentum theory, which is a body-force model combined with RANS to account for two-way coupling with the duct. BET is the blade element theory, which is combined with RANS to calculate separately the flow in an empty duct and turbine performance under such flow (i.e., a simpler version of RANS+BEMT with only one-way coupling).}
\label{table:literature}
\end{table}

To account for complex turbine-duct interaction and more reliably evaluate ducted turbine performance, we must resort to computational fluid dynamics (CFD) simulations or experiments.
Table~\ref{table:literature} lists research efforts that used such approaches. Most studies were conducted for wind turbines, but a few were specific to hydrokinetic turbines.
As shown in the table, various duct shapes have been proposed and tested for wind turbines, with reported efficiencies ranging from 0.41 to 0.85, surpassing the Betz’s limit (0.59).
However, these results must be interpreted in the context of the limitations of the analyses.
The CFD models used for evaluations include steady Reynolds-averaged Navier--Stokes (RANS) and unsteady RANS (URANS) solvers, where the turbine blades are modeled using a blade-resolved or a body-force approach. Within these approaches, steady RANS may have difficulties with flow separation along the duct surface in many designs, as well as in capturing transient wake flow patterns and turbine-duct interactions~\cite{voigt2010aerodynamic,carnevale2013comparative,zhang2020investigation}. In the body-force approach the actuator disk model measures the extracted power using the product of velocity and thrust at the blade section, which usually results in an over-prediction of the efficiency because only a fraction of the computed power can be converted to the actual (rotational) power.

In addition, the definitions of efficiency for ducted turbines are inconsistent in the studies listed in Table~\ref{table:literature}.
The inflow power is defined with respect to either the blade swept area or the maximum projection area of the device (or duct).
The efficiency based on the blade swept area, $C_{P,A_b}$, can be significantly higher than that based on the device area, $C_{P,A_\text{max}}$, but $C_{P,A_b}$ does not provide a fair comparison with the efficiency of an unducted turbine as explained later.
In evaluating and comparing the performance of ducted turbine designs, we must use the same metric, so we convert all power coefficients using the maximum area as the reference---$C_{P,A_\text{max}}$ in Table~\ref{table:literature}. 

Considering the above two points, there are significant caveats in the results listed in Table~\ref{table:literature}.
The highest fidelity simulation in Table~\ref{table:literature} (the blade-resolved URANS approach performed by \citet{Knight2018}) predicts a $C_{P,A_\text{max}}$ of $45\%$, which does not show an advantage of using the duct. The experimental evaluations of \citet{oka2015aerodynamic} may be more credible, but they also suffer from uncertainties, such as measurement errors and proximity of the device to the floor, which causes blockage that affects the measured efficiency~\cite{schluntz2015effect}.
Finally, most of the results obtained for wind turbines do not translate to hydrokinetic turbines.
For example, to sustain higher loads in water, a wind turbine design with a large flange may not be feasible for a hydrokinetic turbine.
Additionally, a hydrokinetic turbine blade requires a lower aspect ratio and larger sectional thickness to sustain the higher loads in water~\cite{gonabadi2021structural}.

Another limitation of the research listed in Table~\ref{table:literature} is that the duct designs were not optimized.
Instead, these designs were generated by human intuition or a grid search in a low-dimensional design space.
The only exception is the design by \citet{aranake2017aerodynamic}, who performed gradient-based optimization to develop a ducted wind turbine design.
However, they used low-fidelity blade element theory to model the ducted turbine performance without adequately taking turbine-duct interaction into account.
An optimal ducted turbine requires numerical optimization that simultaneously considers the blade and duct geometry with detailed shape parametrization.
This is challenging because of the high computational cost of CFD evaluations and the high-dimensional design space.
Another challenge is selecting the appropriate CFD model in the optimization process.
As mentioned earlier, steady RANS is relatively inexpensive but may lead to inaccuracies in predicting the performance of designs where boundary layer separation occurs in the duct. 

In this paper, we perform CFD-based design optimization of a ducted hydrokinetic turbine.
We use 21 parameters to control the shape of the duct (length and multiple sectional radii) and turbine blades (pitch and spanwise twist/chord distributions).
We perform gradient-based optimization with gradients computed by a discrete adjoint method~\cite{he2019design} coupled with steady RANS blade-resolved simulations.
This effort builds on previous design optimizations of unducted wind turbine~\cite{dhert2017aerodynamic,madsen2019multipoint}.
Because of the potential inaccuracies of steady RANS for separated flow, the success of this approach hinges on whether our gradient-based optimization induces a design free of flow separation.
This is, fortunately, indeed the case since designs with flow separation tend to be associated with lower efficiency, even when evaluated by the less accurate RANS solver (given enough grid resolution). 
Our optimized design is re-evaluated by a higher-fidelity URANS blade-resolved solver. The benefits of the duct are demonstrated upon a comparison with the unducted Bahaj turbine, optimized unducted turbine, and our baseline ducted turbines. We follow up with discussions to provide insights on the optimized geometry and the associated flow mechanisms that contribute to improved energy extraction efficiency.

The paper is organized as follows.
In Section~\ref{sec:probstate}, we discuss the problem statement, including the description of the physical problem of turbine energy extraction and the setup of the optimization problem.
Section~\ref{sec:method} introduces methodology in CFD simulations and optimization process.
The results of optimization and higher-fidelity re-evaluation are described in Section~\ref{sec:resultdiscussion}, where we discuss the optimized duct geometry and flow mechanisms.
Finally, conclusions are provided in Section~\ref{sec:conclusion}.  The computations involved in this work are implemented in open-source codes OpenFOAM \cite{jasak2007openfoam} and DAFoam \cite{he2020dafoam}.

\section{Problem Statement}
\label{sec:probstate}

\subsection{Physical Problem}

\begin{figure}[hbt!]

\centering
    \begin{subfigure}[b]{0.35\textwidth}
    \includegraphics[width=\textwidth]{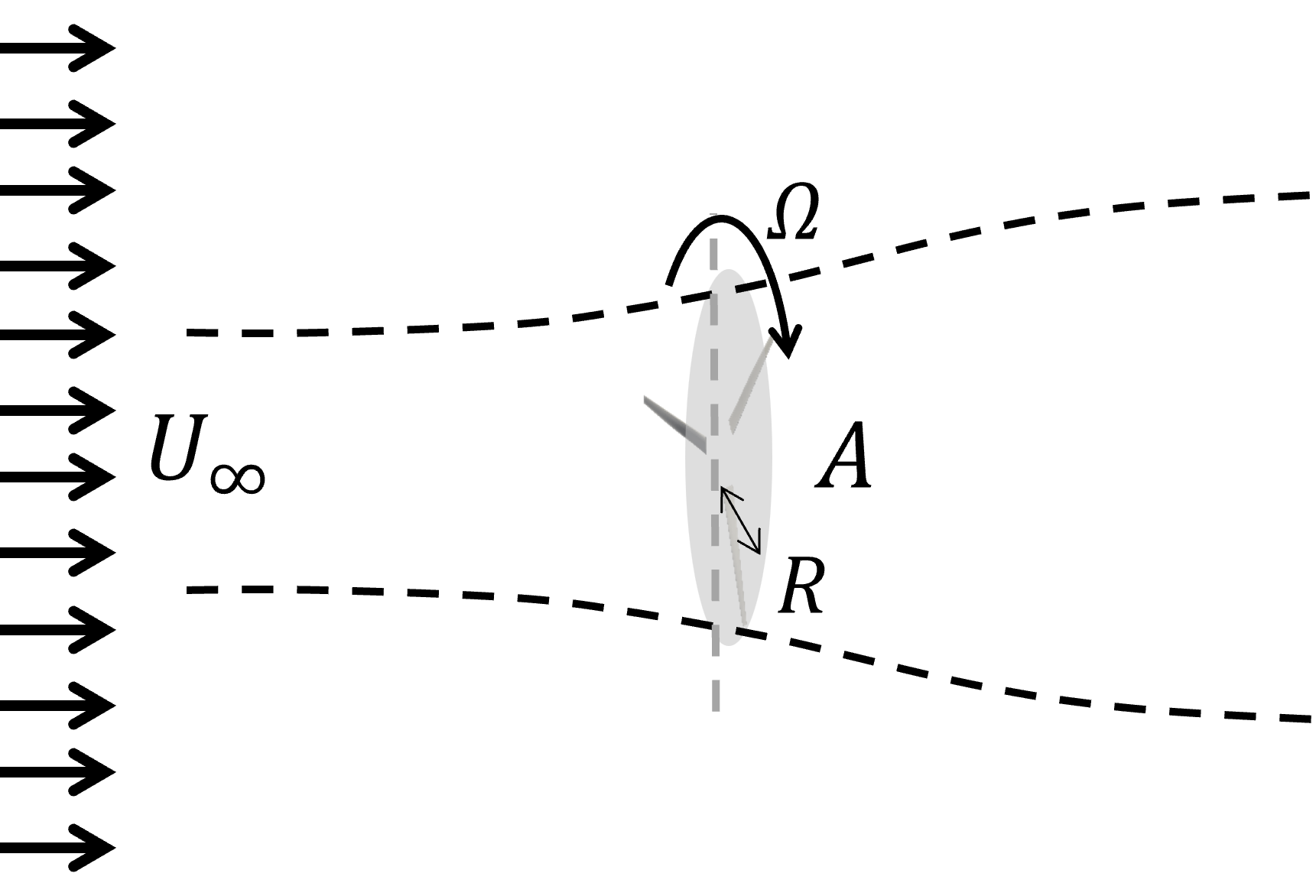}
    \caption{Unducted turbine}
    \end{subfigure}
\hspace{2em}
    \begin{subfigure}[b]{0.35\textwidth}
    \includegraphics[width=\textwidth]{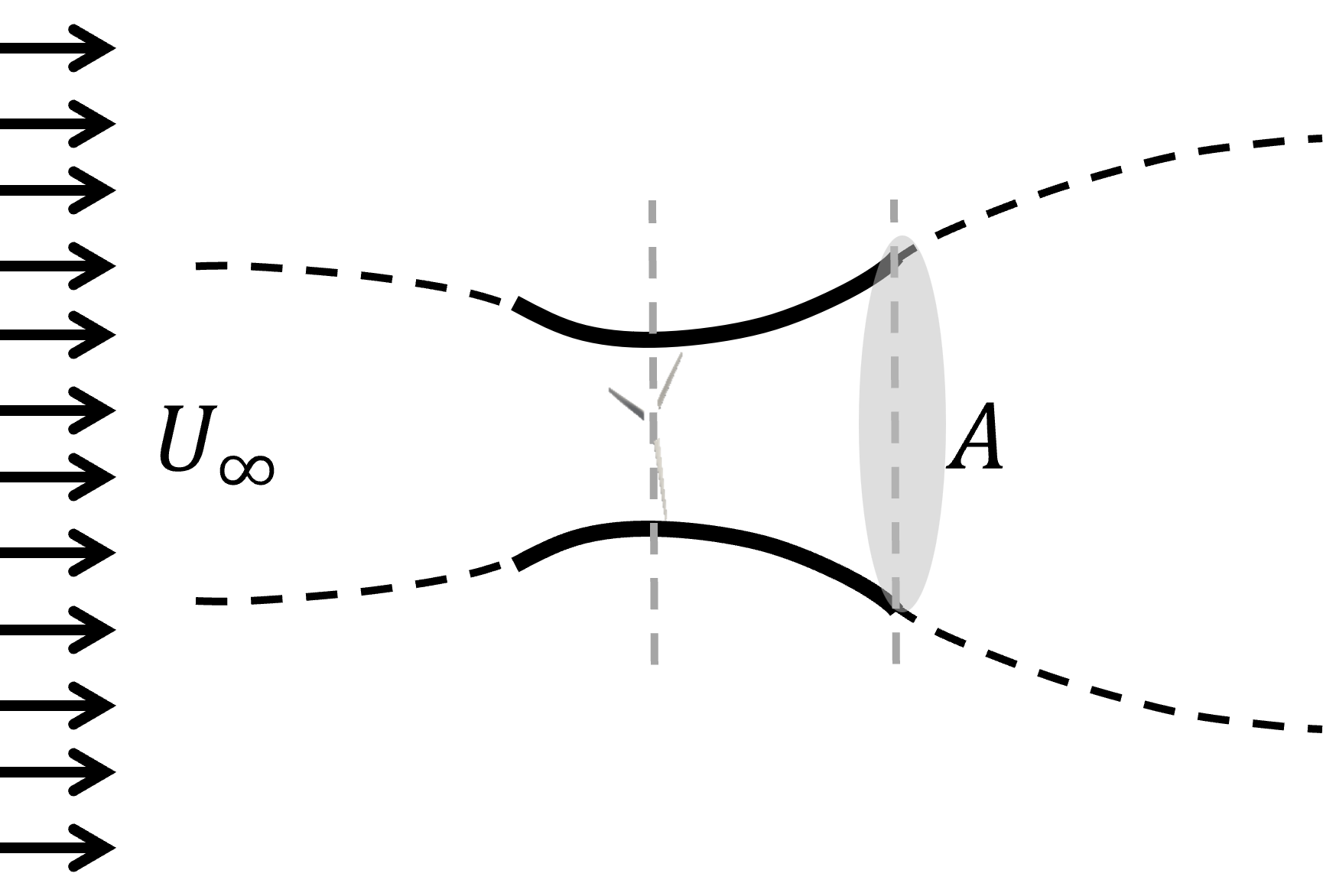}
    \caption{Ducted turbine}
    \end{subfigure}
\caption{Physical problem of (a) an unducted turbine and (b) a ducted turbine, with the same device area $A$, subject to inflow $U_\infty$ in an unbounded domain.}
\label{fig:probstate}
\end{figure}

Consider a turbine operating in a uniform inflow $U_{\infty}$ in an unbounded fluid domain, as shown in Figure~\ref{fig:probstate}.
The turbine converts inflow power (energy) into rotational power, where the effectiveness of this conversion is characterized by the power coefficient,
\begin{equation}
    C_P = \frac{P}{\frac{1}{2}\rho A U_{\infty}^3},
\label{eq_CP}
\end{equation}
where $P$ is the generated rotational power that is given by $P=Q\Omega$ (torque $Q$ times the rotational speed of the blades $\Omega$). $\rho$ is the fluid density and $A$ is some reference area. For an unducted turbine, $A$ can be chosen as either the blade swept area $A_b$ or the maximum projection area of the device $A_\text{max}$, which are identical.
For a ducted turbine, however, using the two values $A_\text{max}$ and $A_b$ as $A$ leads to different $C_P$'s, since $A_\text{max}$ is greater than $A_b$. We argue that $A_\text{max}$ is the appropriate choice for ducted turbines in order to have a fair comparison of their performance with unducted turbines. The reason is that, with $A=A_\text{max}$, we are essentially comparing the generated power when the inflow power is the same for unducted and ducted turbines. On the other hand, using $A_b$ for ducted turbines results in a larger value of $C_P$ (even above 1) that can be misleading when compared to the efficiency of unducted turbines (see examples in~\cite{coiro2016diffuser,tampier2017numerical,song2019numerical}). For the above reasons, we adopt $A=A_\text{max}$ for the evaluation of $C_P$ in this work, and we will hereafter simply write it as $A$, referring to the maximum device area for both ducted and unducted turbines. 

Given a turbine, its efficiency $C_P$ is in general a function of two other non-dimensional parameters, namely the tip-speed ratio ($\lambda$) and Reynolds number $Re$ (based on the diameter of the device), defined as 
\begin{equation}
    \lambda = \frac{\Omega R}{U_{\infty}}, \ \ \ Re=\frac{U_{\infty}D_\text{max}}{\nu}
    \label{eq_TSR},
\end{equation}
where $R$ is the turbine blade radius, $\nu$ is the fluid kinematic viscosity. In this paper, we fix $U_\infty=1.4m/s$, $\nu=1\times10^{-6}m^2/s$, $A=1.853m^2$,  and $D_\text{max}=\sqrt{(4/\pi)A}=1.536m$, leading to $Re\approx2\times10^6$ for both ducted and unducted turbines (see Figure~\ref{fig:probstate}). For Reynolds number of $\mathcal{O}(10^6)$, the flow is considered fully turbulent, and the dependence of $C_P$ on $Re$ in this range is expected to be relatively weak. We will evaluate $C_P$ for a broad range of $\lambda$ at this Reynolds number for both ducted and unducted turbines.

\subsection{Optimization Problem}
\label{subsec:optproblem}
Our objective is to optimize a ducted turbine geometry to maximize its hydrodynamic efficiency $C_P$ at given $U_{\infty}$($=1.4m/s$) and $\Omega$($=17.5rad/s$). This design process is applied to both ducted and unducted turbines for a fair performance comparison. In the following, we present the mathematical optimization problem for a ducted turbine, which is the more sophisticated case. The optimization for an unducted turbine can be conducted similarly but with a simpler setup that does not include the duct parameters and the tip clearance constraint.

\begin{figure}[hbt!]
\centering
    \begin{subfigure}[b]{0.45\textwidth}
    \includegraphics[width=\textwidth]{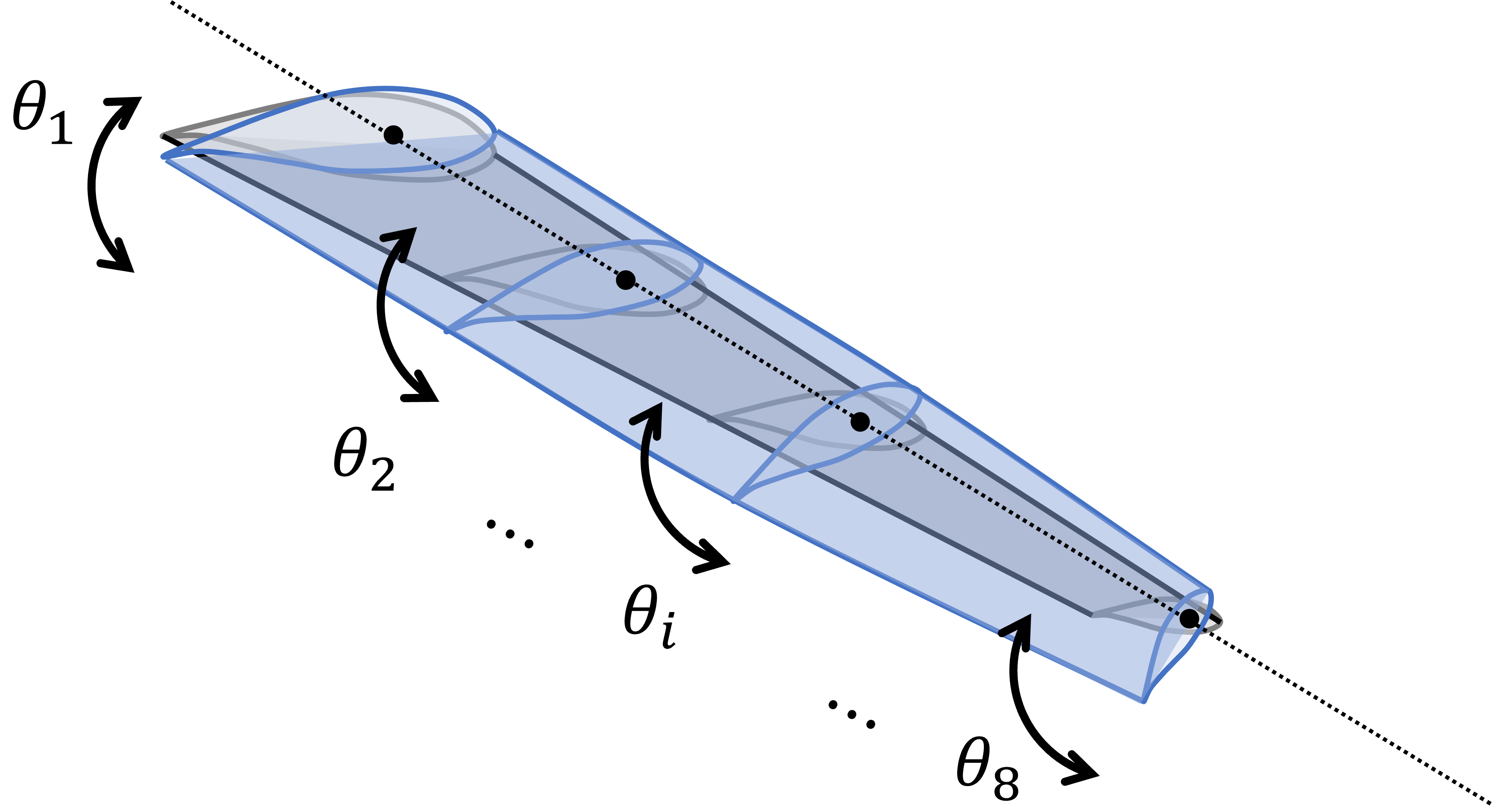}
    \caption{Twist angles}
    \label{fig:bladeDV1}
    \end{subfigure}
\hspace{0.5em}
    \begin{subfigure}[b]{0.45\textwidth}
    \includegraphics[width=\textwidth]{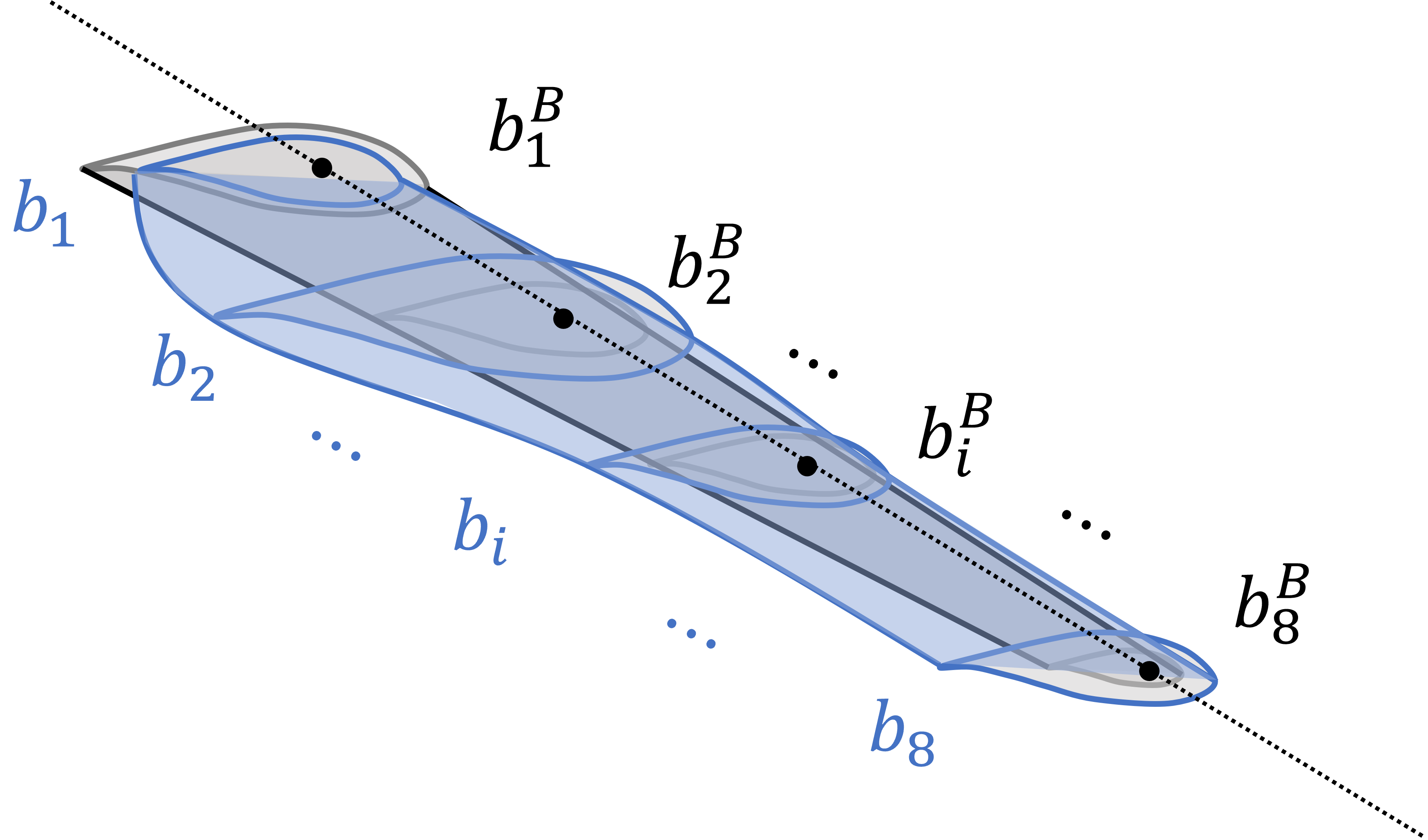}
    \caption{Blade section scales}
    \label{fig:bladeDV2}
    \end{subfigure}
    \caption{Design variables of blades. Gray and blue colors represent original and modified designs, respectively. (a) Blade twist angles ($\{\theta_i\}_{i=1}^8$). (b) Blade section areas ($\{b_i\}_{i=1}^8$), where we also include the baseline $\{b_i^B\}_{i=1}^8$. Blade sections can be contracted ($0< b_i/b^B_i< 1$) or expanded ($b_i/b^B_i > 1$).}
    \label{fig:bladeDV}
\end{figure}

The constrained optimization problem for a ducted turbine can be stated as
\begin{linenomath}
\begin{subequations}
\label{eq:optproblem}
\begin{align}
    \text{maximize}
        & \quad C_P  \label{eq_optobjective}\\
    \text{by varying}
        & \quad -30^{\circ} \leq  \{ \theta_i \}_{i=1}^{8} \leq 30^{\circ}, \label{eq_dv1} \\
        & \quad 0.8 \leq  \Bigl\{ \frac{b_i}{b^B_i} \Bigl\}_{i=1}^{8} \leq 1.2, \label{eq_dv2} \\
        & \quad 0 \leq d_3 \leq \{d_j\}_{j=1,2,4} \leq D_\text{exit}, \label{eq_dv3} \\
        & \quad 0.3 \leq \frac{l}{l^B} \leq 1.5,  \label{eq_dv4} \\
    \text{subject to}
        & \quad  \frac{2R}{d_3} = 0.91,  \label{eq_optconstraint}
\end{align}
\end{subequations}
\end{linenomath}
where $\{ \theta_i \}_{i=1}^{8}$ are the twist angles at 8 sections of the blade, controlling the blade root pitch and twist profile as shown in Figure~\ref{fig:bladeDV1}. The cross-sectional areas of 8 sections of the blade, normalized with respect to their baseline, are denoted as $\{ b_i/b_i^B \}_{i=1}^{8}$. Modifying these variables leads to a change in the size of the blade section, but the sectional (foil) shape remains unchanged. Thus, $b_i/b^B_i$ gives the scaling factor for each section, as shown in Figure~\ref{fig:bladeDV2}.

\begin{figure}[hbt!]
    
\centering
\begin{subfigure}[c]{\textwidth}
    \includegraphics[width=\textwidth]{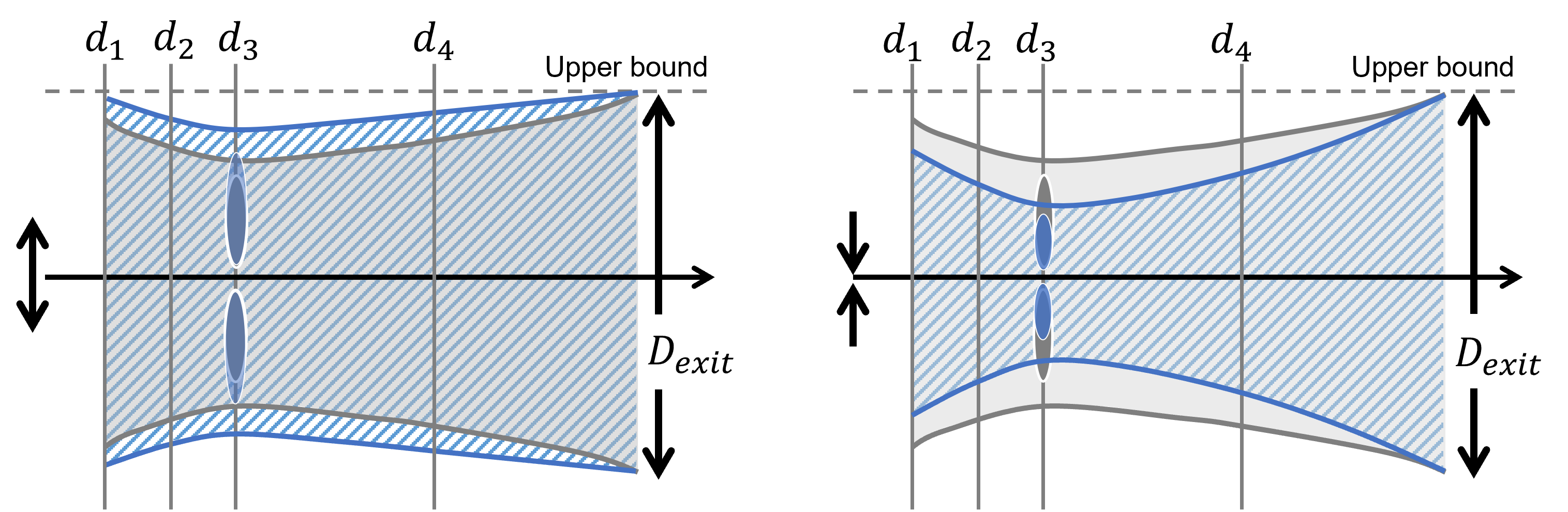}
    \caption{Duct sectional radii change. Expansion and contraction}
    \label{fig:ductDV1}
\end{subfigure}
\begin{subfigure}[c]{\textwidth}
    \includegraphics[width=\textwidth]{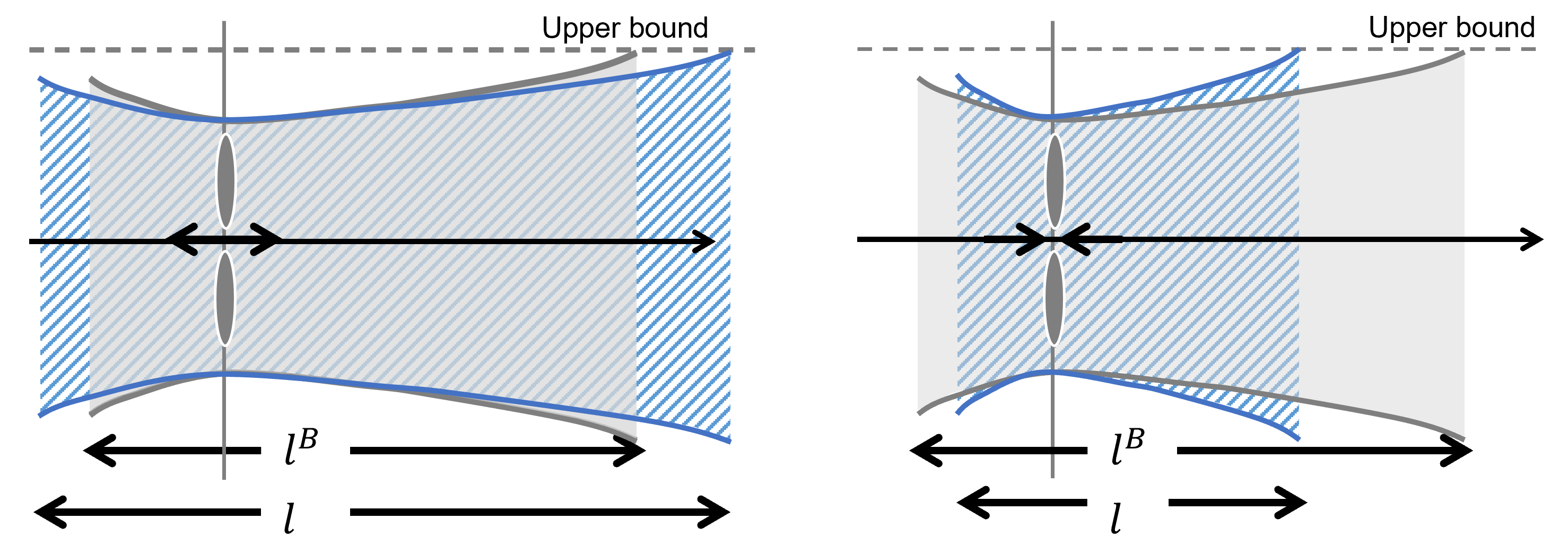}
    \caption{Duct length change. Elongation and shortening}
    \label{fig:ductDV2}
\end{subfigure}

\caption{Design variables of a duct. Gray and blue colors represent original and modified designs, respectively. (a) Duct radial expansion (left) and contraction (right), controlled by sectional diameter variables $\{d_i\}_{i=1}^4$, which are bounded above by $D_\text{exit}$ and below by the variable $d_3$ at the throat. The blade radius $R$ is scaled to maintain the tip gap ratio with $2R/d_3 = 0.91$. (b) Duct elongation (left) and shortening (right), controlled by the design variable $l$, with respectively $l/l^B \geq 1$ and $0 \leq l/l^B \leq 1$.} 
\label{fig:ductDV}
\end{figure}

The variables $\{d_j\}_{j=1}^4$ are the diameters at 4 sections along the duct as illustrated in Figure~\ref{fig:ductDV1}, with $d_3$ being the throat section, where the rotor is installed. This section is located at 26.4\% of the duct length, following a baseline design of the duct~\cite{Knight2018}. The bound~\eqref{eq_dv3} ensures that $d_3$ is always located at the throat in the optimization and that all duct diameters do not exceed the exit diameter $D_\text{exit}=D_\text{max}=\sqrt{(4/\pi)A}$, as depicted in Figure~\ref{fig:ductDV1}. A large exit area reduces flow velocity at the exit through the streamtube expansion, which in turn increases the flow momentum extraction at the blades. Our setup prevents the optimizer from unrealistically increasing the size of the exit section and ensures a fair comparison between different designs, as discussed in Sec.~\ref{sec:intro}.
The variable $l$, representing the duct length (with $l^B$ the baseline value), governs the scaling of the duct with respect to the fixed point at the throat (Figure~\ref{fig:ductDV2}). The constraint~\eqref{eq_optconstraint} keeps the tip gap ratio as a constant of 9\% throughout the optimization process, consistent with the baseline design. It leads to the blade radius $R$ changing with the variation of throat diameter $d_3$.

The design variable bounds are set up in a trial-and-error manner to make sure that the optimized variables do not reach the bounds on the optimized design. The full turbine geometry morphs smoothly throughout the optimization. This continuous morphing is ensured through the Free-Form Deformation method which will be discussed later in Section~\ref{subsubsec:ffd}.

\begin{figure}[hbt!]
\centering
\begin{subfigure}[c]{0.8\textwidth}
    \includegraphics[width=\textwidth]{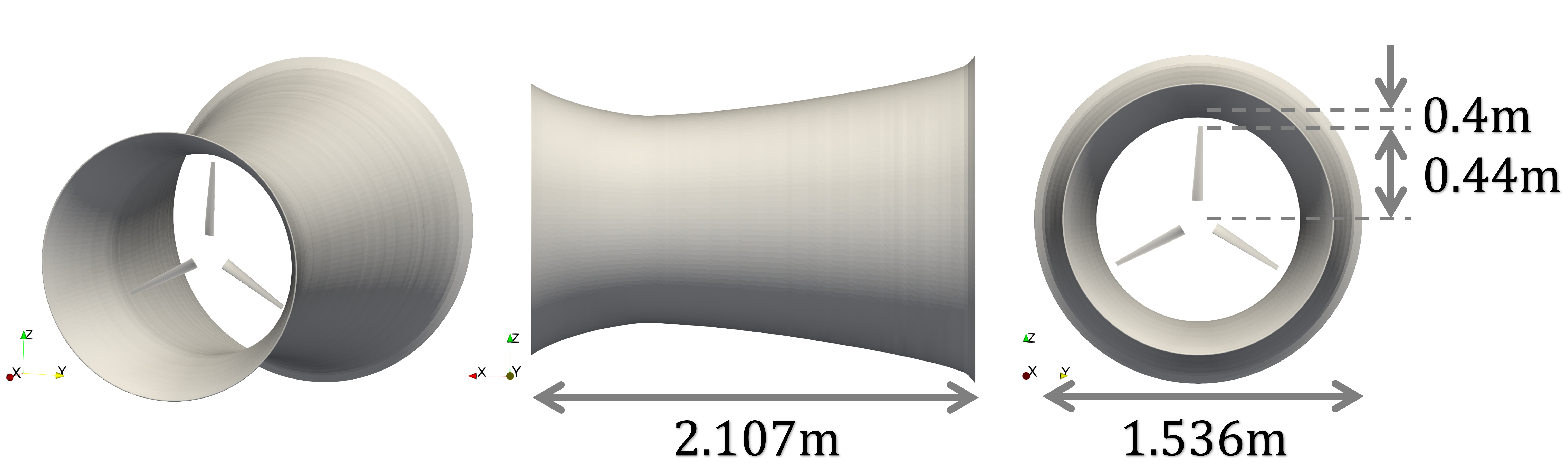}
    \caption{Different views of baseline ducted turbine}
\end{subfigure}
\begin{subfigure}[c]{0.7\textwidth}
    \includegraphics[width=\textwidth]{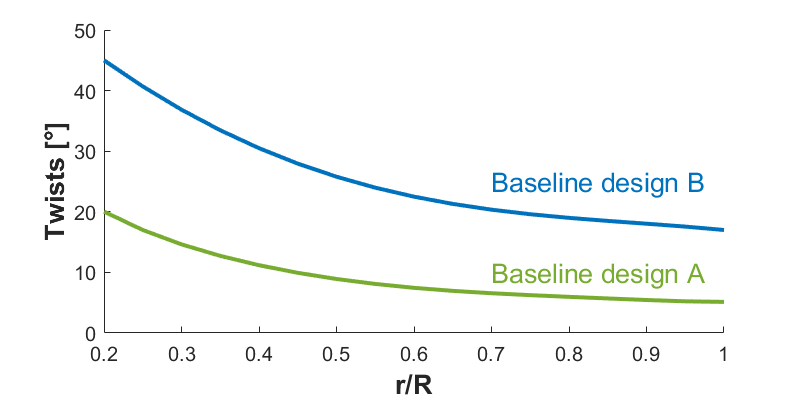}
    \caption{Twist distributions of baseline designs A and B}
\end{subfigure}
    \caption{Baseline designs of a ducted turbine. (a) Different views of baseline ducted turbine; (b) Twist distributions of baseline designs A and B.}
    \label{fig:baseline}
\end{figure}

Since we use a gradient-based optimization method~\eqref{eq:optproblem}, local optima potentially exist in the design space. Hence, we adopt the multistart strategy, using two different baseline designs (hereafter named baseline design A and B) with drastically different performances.
Both baseline designs adopt the same thin-wall curved-shaped duct as in~\cite{Knight2018}. The two baseline designs differ in the blade geometry (see Figure~\ref{fig:baseline}).
Design A adopts the original Bahaj model with a $0.44m$ radius and a $20^\circ$ root pitch.
Design B adopts the Bahaj model with a $0.44m$ radius, a $45^\circ$ root pitch, and a modified twist profile as in~\cite{Knight2018}. This modified twist profile is obtained by matching the local angle of attack of each blade section in the duct to that of the unducted turbine counterpart through an iterative procedure. 
When evaluating with the unsteady RANS solver, baseline designs A and B yield $C_P=28\%$ and $C_P=45\%$, respectively, at $\lambda=5.5$. The hub is not included in the model to simplify the geometry parametrization using the Free-Form Deformation method. 
The optimization problem is summarized in Table~\ref{table:optSetup}.

\begin{table}[h!]
\centering
\begin{adjustbox}{width=\columnwidth,center}
\begin{tabular}{llcc}
    
    \multicolumn{2}{l}{\multirow{2}{*}{\textbf{}}} 
                 & \multirow{2}{*}{\textbf{\textbf{Optimization A (design A)}}} 
                 & \multirow{2}{*}{\textbf{\textbf{Optimization B (design B)}}}\\ 
    \multicolumn{2}{l}{} & &\\
    \hline

    \multicolumn{2}{l}{\multirow{2}{*}{\textbf{Objective}}} & \multicolumn{2}{c}{\multirow{2}{*}{Maximize $C_{P}$ at fixed $U_{\infty}$ = 1.4m and $\Omega$ = 17.5 rad/s}} \\
    \multicolumn{2}{l}{} & \multicolumn{2}{l}{}\\

    \multirow{4}{*}{\textbf{Baseline Design}} 
          & \multirow{2}{*}{\textbf{Blade}} & \multirow{2}{*}{Original-twist Bahaj~\cite{bahaj2007power}} & \multirow{2}{*}{Twist-modified Bahaj~\cite{Knight2018}} \\
     & & & \\
    
     & \multirow{2}{*}{\textbf{Duct}} & \multicolumn{2}{c}{\multirow{2}{*}{Thin-walled duct from \citet{Knight2018}}} \\
     & & \multicolumn{2}{l}{} \\

    \multirow{4}{*}{\textbf{Design Variables}} 
          & \multirow{2}{*}{\textbf{Blade}} & \multicolumn{2}{c}{\multirow{2}{*}{$\{ \theta_i, b_i \}_{i=1}^{8}$ (16 vars)}} \\
     & & \multicolumn{2}{l}{} \\
    
     & \multirow{2}{*}{\textbf{Duct}} & \multicolumn{2}{c}{\multirow{2}{*}{$\{d_j\}_{j=1}^{4}, l$ (5 vars)}} \\
     & & \multicolumn{2}{l}{} \\

    \multicolumn{2}{l}{\multirow{2}{*}{\textbf{Bounds and Constraints}}} & \multicolumn{2}{c}{ \multirow{2}{*}{~\eqref{eq_dv1} to ~\eqref{eq_optconstraint}}} \\
    \multicolumn{2}{l}{} & \multicolumn{2}{l}{}\\
    \hline \hline
\end{tabular}
\end{adjustbox}
\caption{Setup of optimizations A and B}
\label{table:optSetup}
\end{table}

\section{Methodology}
\label{sec:method}
\begin{figure}[hbt!]
    \centering
    \includegraphics[width=1.0\textwidth]{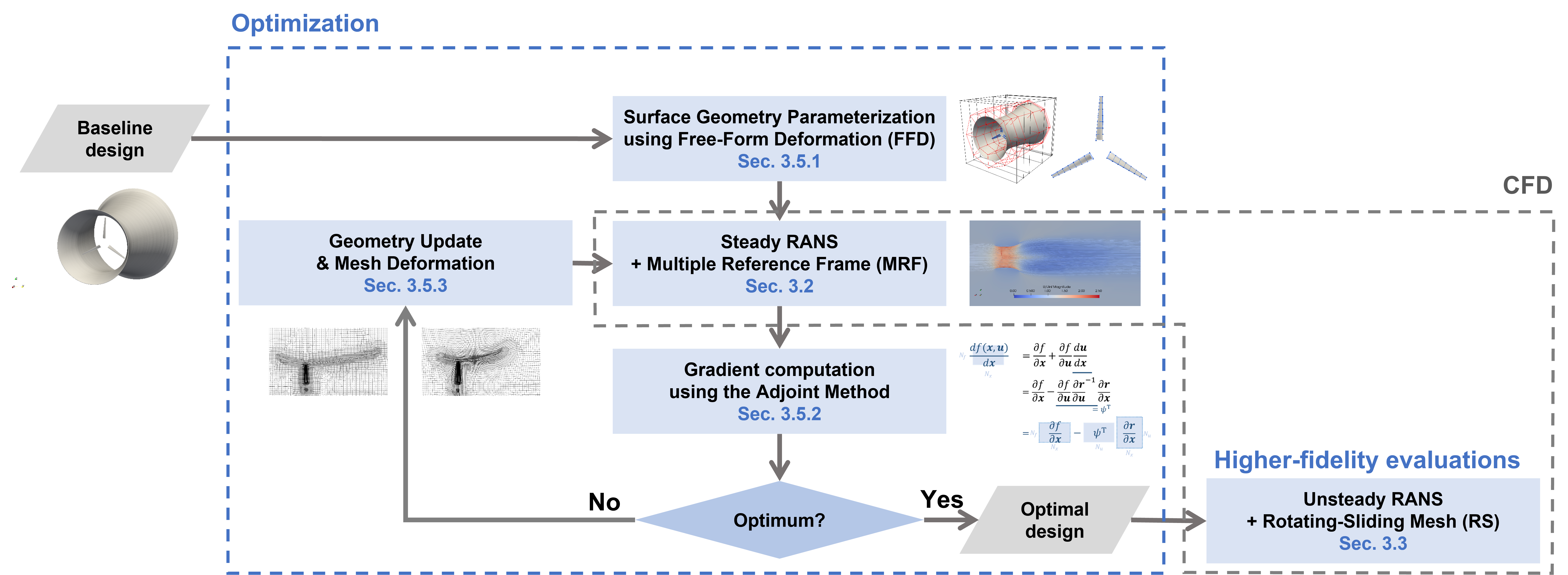}
    \caption{Diagram of the overall process, which includes optimization and re-evaluation of the optimal designs.}
    \label{fig:optProcedure}
\end{figure}

This section describes the methodology in optimization and CFD evaluations. An overall flowchart is shown in Figure~\ref{fig:optProcedure}.
The whole process involves the optimization and the re-evaluation of the optimized design using higher-fidelity simulations. 
The optimization and high-fidelity re-evaluation use DAFoam~\cite{he2020dafoam} and OpenFOAM~\cite{jasak2007openfoam}, respectively. In what follows, we will describe each component of the methodology in subsections, as outlined in Figure~\ref{fig:optProcedure}. To provide a self-contained but easy-to-follow paper for readers, we put additional details in the appendix and keep the main paper as concise as possible. We start from CFD models involved in both the optimization and re-evaluations and then follow up with other components in the optimization framework.

\subsection{CFD models}
\label{subsec:cfdmodels}
The governing equations for the flow field around the turbine (Figure~\ref{fig:probstate}) are the Navier--Stokes equations
\begin{subequations}
\label{eq_NavierStokes}
\begin{numcases}{}
        \nabla\cdot \bm{U} = 0 \label{eq_NSmass},\\
        \frac{\partial\bm{U}}{\partial t}+\nabla\cdot (\bm{U}\bm{U}) = -\frac{1}{\rho}\nabla p + \nabla\cdot(\nu\nabla\bm{U}),
        \label{eq_NSmomentum}
\end{numcases}
\end{subequations}
where $\bm{U}$ is the flow velocity and $p$ is the pressure.
We apply velocity inlet and pressure outlet boundary conditions, and no-slip boundary conditions on the blade and duct surfaces.

We consider the Reynolds-averaged Navier--Stokes (RANS) equations with grids only resolving the averaged components of the flow. One can apply the Reynolds decomposition $\bm{U} = \overline{\bm{U}}+\bm{u}'$ (and the same for pressure) to Eq.~\eqref{eq_NavierStokes}.
$\overline{\bm{U}}$ denotes the averaged velocity in a time window or by an ensemble and $\bm{u}'$ represents the zero-mean turbulent fluctuation.
This leads to the unsteady RANS equation:
\begin{equation}
    \begin{cases}
        \nabla\cdot\overline{\bm{U}} = 0, \\
        \frac{\partial\overline{\bm{U}}}{\partial t}+\nabla\cdot (\overline{\bm{U}}\:\overline{\bm{U}}) = -\frac{1}{\rho}\nabla \overline{p} + \nabla\cdot(\nu\nabla\overline{\bm{U}})- \nabla\cdot \overline{\bm{u}'\bm{u}'},
    \end{cases}
\label{eq_RANS}
\end{equation}
where $\overline{\bm{u}'\bm{u}'}$ is the Reynolds stresses that need to be approximated by turbulence models. In this work, we use the $k-\omega$ SST turbulence model together with the automatic near-wall treatment (see~\ref{sec:kwsst:appendix} for details on both). 

The rotating blades are handled in simulations by two blade-resolved approaches: the Multiple Reference Frames (MRF) and the rotating-sliding mesh approach (RS). The former is used for steady RANS solutions (i.e., a solution with time-derivative terms set to zero) with multiple different reference frames, while the latter is used directly in the unsteady solution of Eq.~\eqref{eq_RANS}. 
The MRF is used in optimization.
The RS is defined as the higher-fidelity approach and used for optimized result re-evaluations (see Figure~\ref{fig:optProcedure}).
We include a detailed introduction of the two approaches in the following sections.

\subsection{Multiple Reference Frames (MRF) Method}
\label{subsec:mrf}
\begin{figure}[hbt!]
    \centering
    \includegraphics[width=0.6\textwidth]{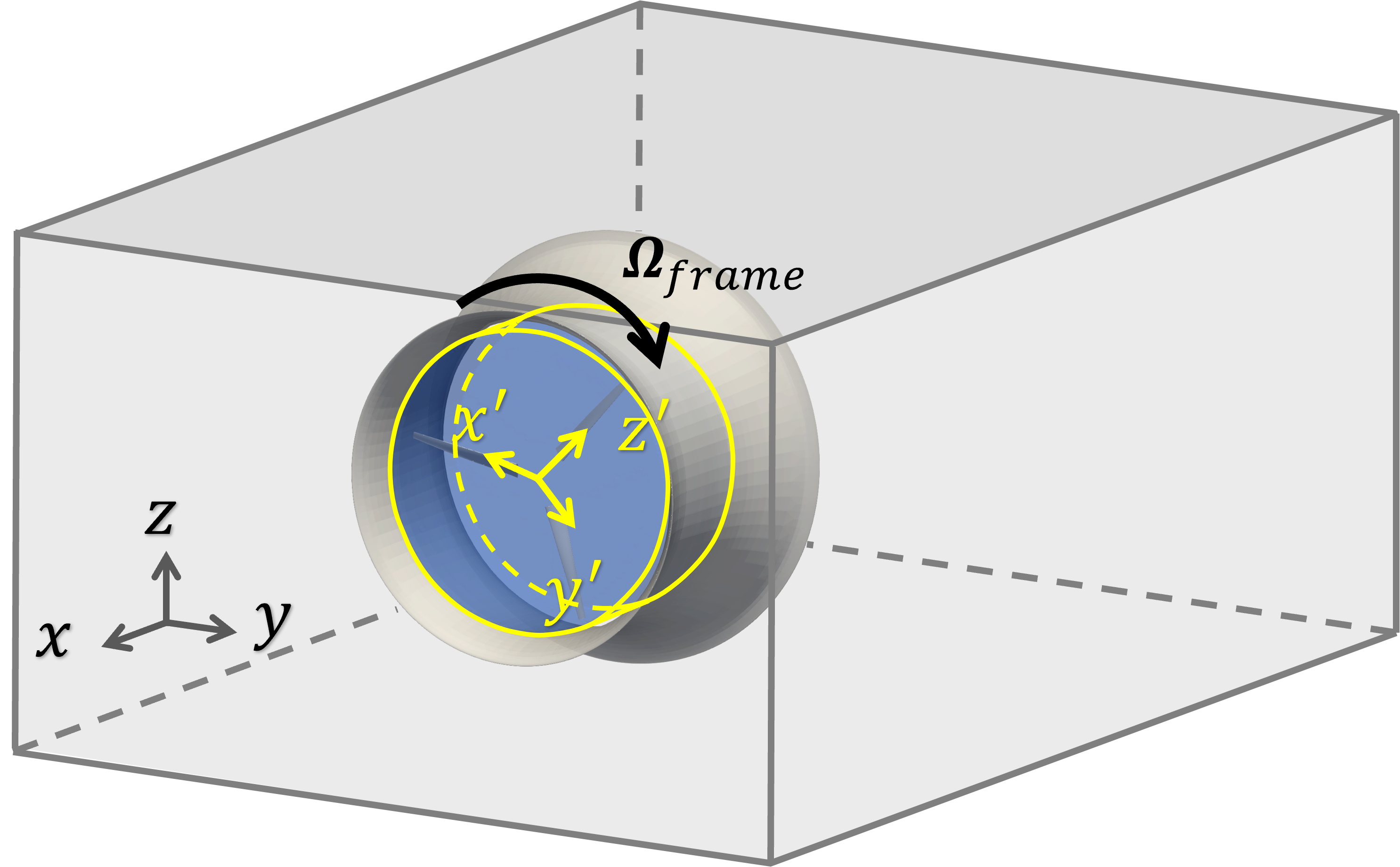}
    \caption{MRF method illustration. The computational domain is split into stationary and rotating (shaded blue) regions. The inertial and rotating reference frames are denoted by gray and yellow axes, respectively.}
    \label{fig:MRF:tworegions}
\end{figure}

The MRF method is an efficient method for modeling turbomachinery flow. In the MRF method, the computational mesh stays stationary, and the rotational effect is handled through a rotational reference frame. In particular, the fluid domain is separated into two regions: a rotational region surrounding the turbine blades with a blade-fixed reference frame, and the remaining stationary region with an inertial reference frame, as shown in Figure~\ref{fig:MRF:tworegions}. In both regions, the flow is considered steady with respect to the corresponding reference frame, so only steady RANS equations need to be solved. 

To be more specific, in the rotational region, the blades are stationary and experience a steady inflow. The flow velocity in the blade-fixed reference frame can be expressed by
\begin{equation}
    \bm{U}_R=\bm{U}-\bm{\Omega}\times\bm{r},
\label{eq:MRF:relativeVel}
\end{equation}
where $\bm{U}$ is the velocity in the inertial reference frame, $\bm{\Omega}$ is the rotation vector of the turbine blades, and $\bm{r}$ is the distance vector from the axis of rotation to the point of interest (position vector). The steady RANS equations in the rotational region need to be established with the blade-fixed reference frame, which requires further formulations of both Eq.~\eqref{eq_RANS} and the $k-\omega$ SST model equations. Although the implementation in OpenFOAM/DAFoam solves this complete set of equations, here we only present the rotational-region formulation regarding Eq.~\eqref{eq_NavierStokes} to provide the key insights of the method. 

Combining Eq.~\eqref{eq_NavierStokes} and Eq.~\eqref{eq:MRF:relativeVel}, we obtain (see~\ref{sec:mrf:appendix} for a detailed derivation)
\begin{equation}
    \begin{cases}
        \nabla\cdot \bm{U} = 0, \\
        \frac{\partial\bm{U}_R}{\partial t} + \nabla\cdot (\bm{U}_R\bm{U}) = -\frac{1}{\rho}\nabla p + \nabla\cdot(\nu\nabla\bm{U}) - \bm{\Omega}\times\bm{U}.
    \end{cases}
\label{eq_MRFinrotation}
\end{equation}

The steady equations solved in the rotational region are Eq.~\eqref{eq_MRFinrotation}, with $\partial\bm{U}_R/\partial t=0$. Therefore, in the MRF method, steady versions of Eq.~\eqref{eq_NavierStokes} and Eq.~\eqref{eq_MRFinrotation} are solved in stationary and rotating regions. Solving these steady equations can be done using the SIMPLE algorithm~\cite{ferziger2002computational} implemented as simpleFoam in OpenFOAM. 

Although the RANS-MRF method provides an efficient numerical solution for the turbine problem (i.e., only two steady RANS equations need to be solved), its accuracy can be compromised because of two issues.
First, the rotational region and stationary region are usually chosen in a subjective manner. There is no guarantee that the rotational region covers all the flow features resulting from the rotating and discrete blades. Any mismatch between the choice of the region and the nature of the flow can lead to errors at the interface and thus in the final results.
Secondly, for many designs, the flow can be unsteady in nature, especially when flow separation occurs from the duct and/or blade surfaces. Assuming a steady state solution, as in the RANS-MRF method, can lead to significant errors for this type of unsteady flow. As a result, the RANS-MRF method is considered a lower-fidelity model in the context of this paper.

\subsection{Rotating-sliding mesh approach}
\label{subsec:ami}
\begin{figure}[hbt!]
    \centering
    \includegraphics[width=0.3\textwidth]{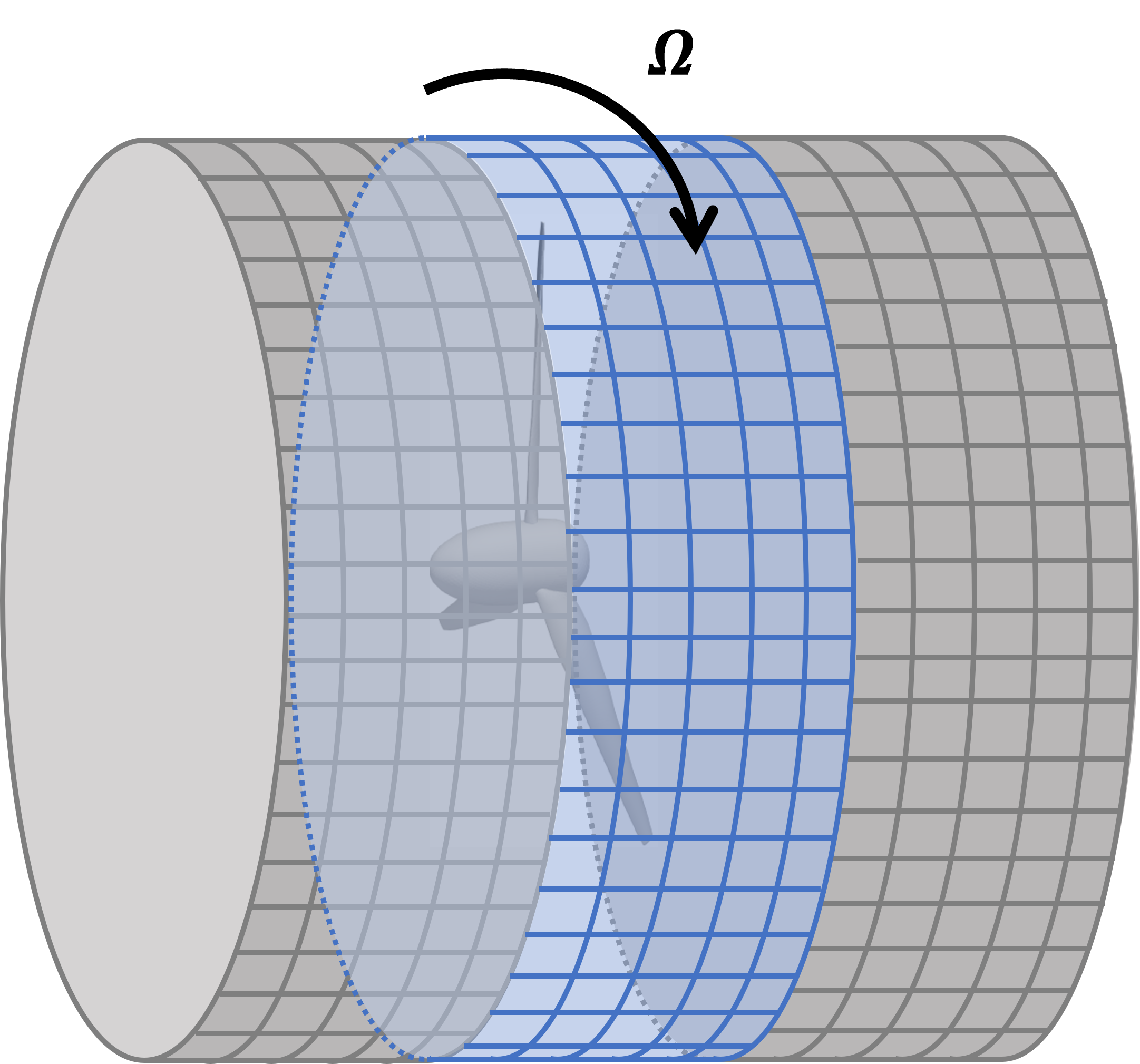}
    \caption{Rotating (blue) and stationary (gray) mesh, with non-conformal mesh interfaces.}
    \label{fig:AMI:nonconform}
\end{figure}

The rotating-sliding mesh (RS) allows for the direct simulation of the unsteady RANS (URANS) equations (Eq.~\eqref{eq_RANS}) with mesh domains that exhibit relative motion. This is needed for modeling rotating geometries. The underlying idea of this method is to allow a region of the computational mesh to rotate with the turbine blades, as illustrated in Figure~\ref{fig:AMI:nonconform}. The rotating mesh also creates a technical problem that the mesh at the rotating/non-rotating interface becomes non-conformal, i.e., the nodes at two sides of the interface do not match up. The data transfer across the interface, therefore, needs to be handled by a special interpolation method involving a ``supermesh''~\cite{farrell2011conservative}, as described in~\ref{sec:ami:appendix}.

Coupling URANS with the RS is, in principle, much more accurate than the RANS-MRF method as it captures the unsteady nature of the flow. This can be critical in simulating flow around a ducted turbine because the possible flow separation from the duct surface can be captured better. However, in the URANS-RS, a small time step is needed to resolve the blade rotation, so a long simulation time is required for the solution to reach a quasi-steady state. 
The computational cost of the URANS-RS is hence much higher than that of the RANS-MRF method. In this paper, the URANS-RS is considered a higher-fidelity method and is only used in re-evaluating the performance of optimized designs. 

\subsection{Mesh Configuration}
\label{subsec:meshconfig}

\begin{figure}[hbt!]
\centering
    \begin{subfigure}[b]{0.28\textwidth}
    \includegraphics[width=\textwidth]{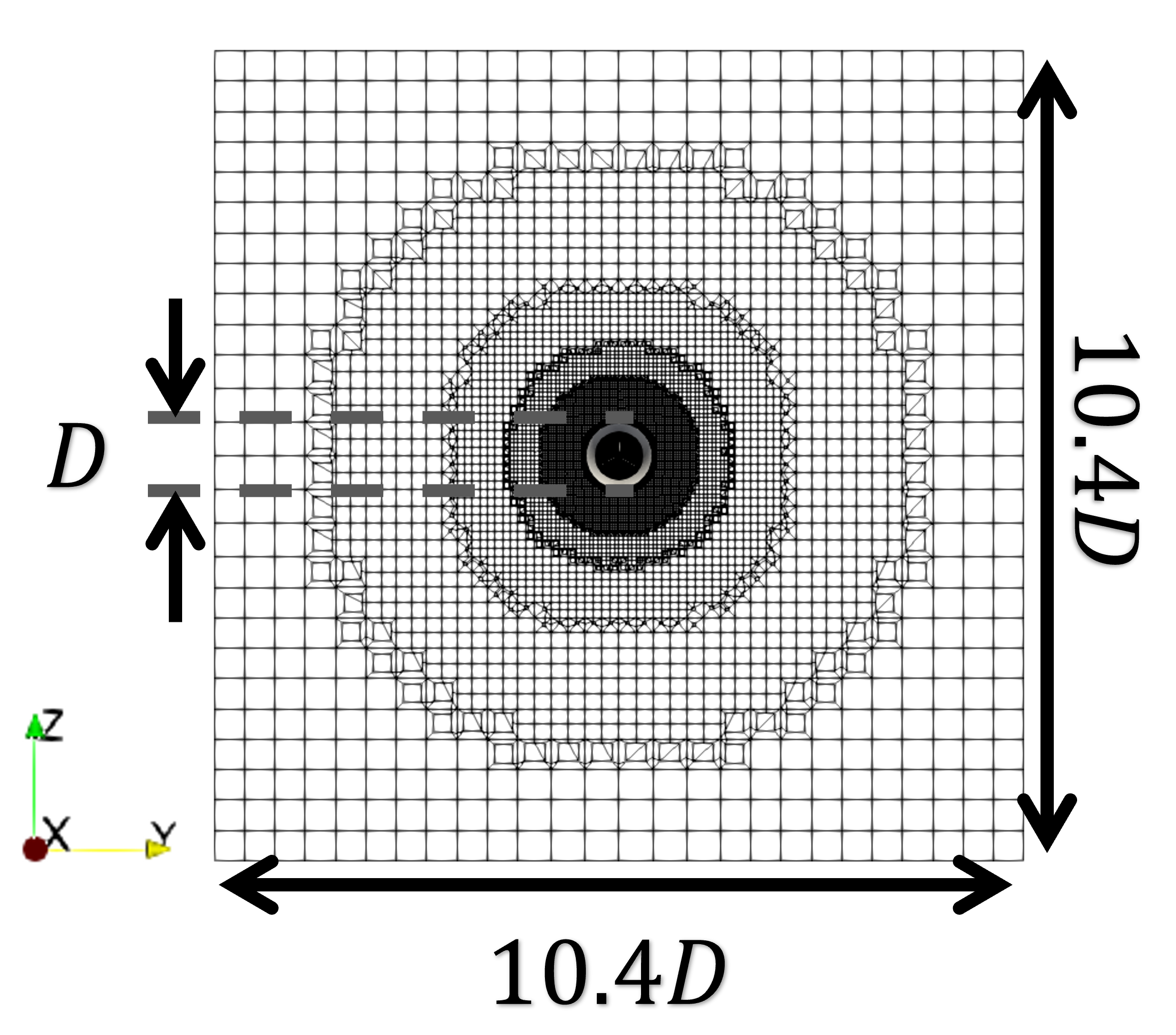}
    \caption{Front view}
    \end{subfigure}
\hspace{0.5em}
    \begin{subfigure}[b]{0.67\textwidth}
    \includegraphics[width=\textwidth]{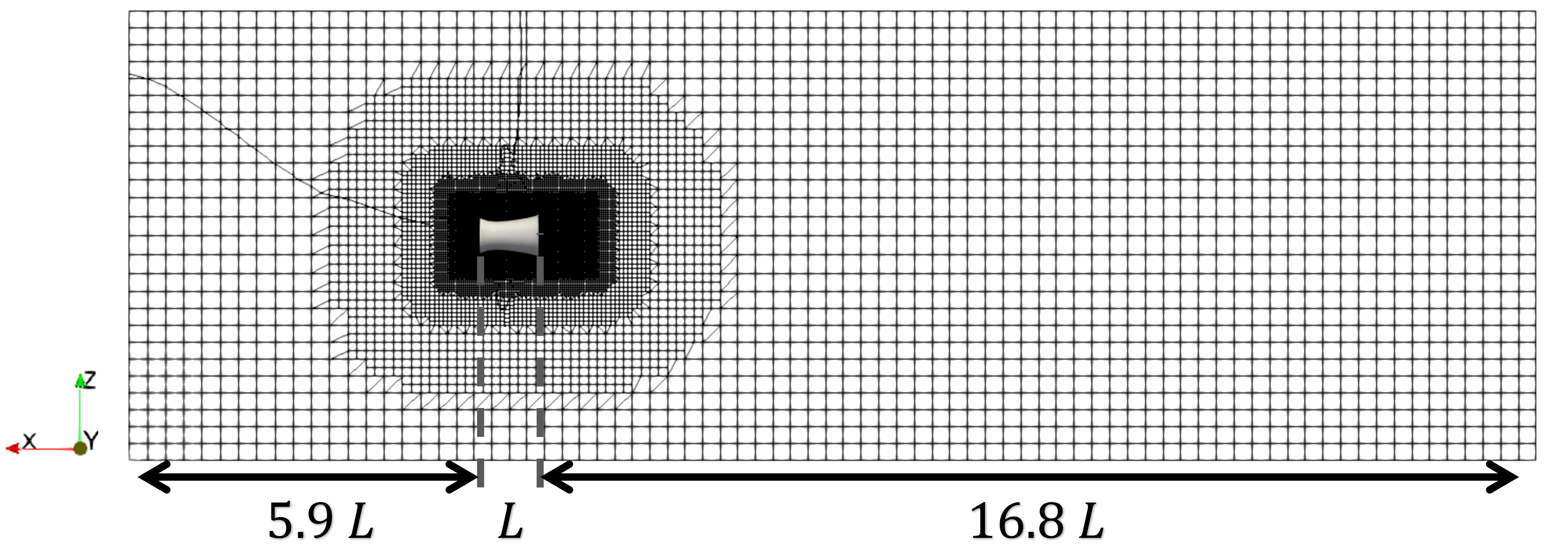}
    \caption{Side view}
    \end{subfigure}
    \begin{subfigure}[b]{0.6\textwidth}
    \includegraphics[width=\textwidth]{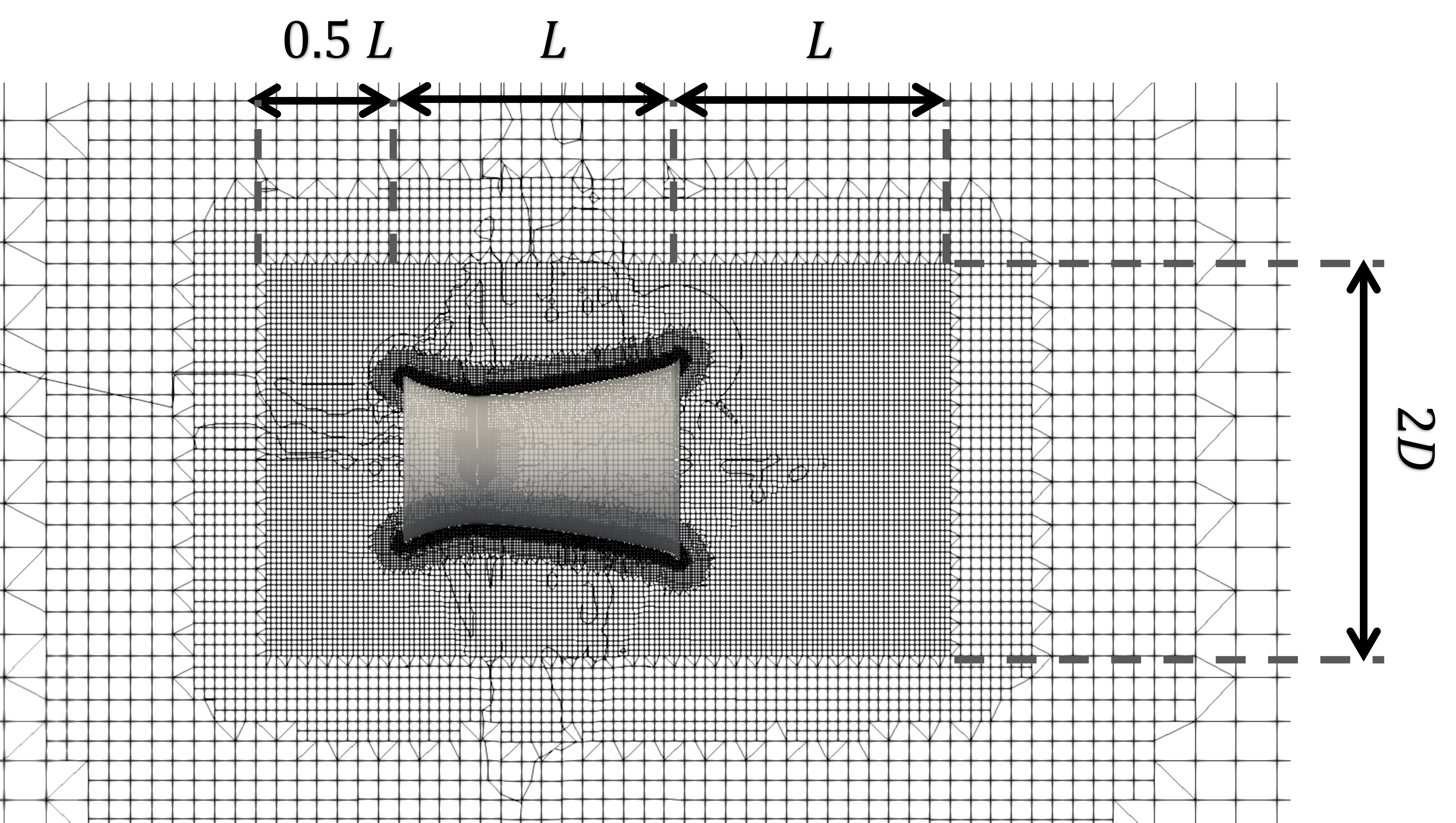}
    \caption{Zoomed-in side view}
    \end{subfigure}
\caption{Mesh configuration for a ducted turbine, with (a) front view, (b) side view, and (c) zoomed-in side view to show the refinement region.}
\label{fig:meshConfig}
\end{figure}

The unstructured computational mesh is generated using the OpenFOAM meshing tool \emph{snappyHexmesh}. A mesh overview is shown in Figure~\ref{fig:meshConfig}. The size of the computational domain is $10.4D\times10.4D\times23.7L$, where $D=\sqrt{(4/\pi)A}=1.536m$ is the maximum diameter of the duct, and $L=2.107m$ is the length of the duct that is taken from the baseline design. This domain size is sufficient to avoid a blockage effect upon tests.

To model flow near the turbine and immediately downstream with higher accuracy, we use a refinement region of $2D\times2D\times2.5L$ around the turbine. Within the refinement region, we first apply a level-4 refinement, i.e., each cell of the original mesh is divided into $(2)^4$ cells in each direction. Then we add around the duct and blade surfaces prism layers that contain further continuously refined cells toward the surfaces (2 and 3 layers are used for the former and latter, both with the expansion ratio of 1.1). The prism layer provides better resolution for boundary layers and is critical for us to obtain well-convergent results in the grid sensitivity study shown later in Section~\ref{sec:resultdiscussion}. In this work, we use three grid resolutions, M0, M1, and M2, with an increasing number of cells, i.e., further refinement from M0 to M2. The coarsest grid M0 is used in the RANS-MRF in the optimization process and has 2-3 million cells (the exact number depends on turbine geometry and re-mesh procedure in optimization). In M1 and M2, the full mesh region (including background mesh and refinement region) is uniformly refined in each direction by the factors of about $1.3$ and $1.6$, respectively, leading to 4-5 million cells for M1 and 7-8 million cells for M2. 
All grids M0, M1, and M2 are used in the URANS-RS re-evaluation, including the grid sensitivity study.  

\subsection{Optimization}
\label{subsec:optMethod}

We use Sequential Quadratic Programming (SQP)~\cite{martins2021engineering}, implemented in SNOPT~\cite{gill2005snopt}, to solve the optimization problem.
One of the challenging tasks is to obtain the gradient of the objective function $\nabla C_P$ with respect to all design variables. 
Sections~\ref{subsubsec:ffd} and~\ref{subsubsec:ajoint} discuss respectively two components (see Figure~\ref{fig:optProcedure}) involved in the gradient computation: 
(1) geometry parametrization and mesh deformation;
(2) adjoint method.
The gradient information is then used in the SQP algorithm to obtain the next design points. We iterate the procedures until satisfying convergence criteria or further design improvements are not achievable. 

\subsubsection{Geometry parametrization via FFD method}
\label{subsubsec:ffd}

\begin{figure}[hbt!]
\centering
    \begin{subfigure}[b]{0.4\textwidth}
    \includegraphics[width=\textwidth]{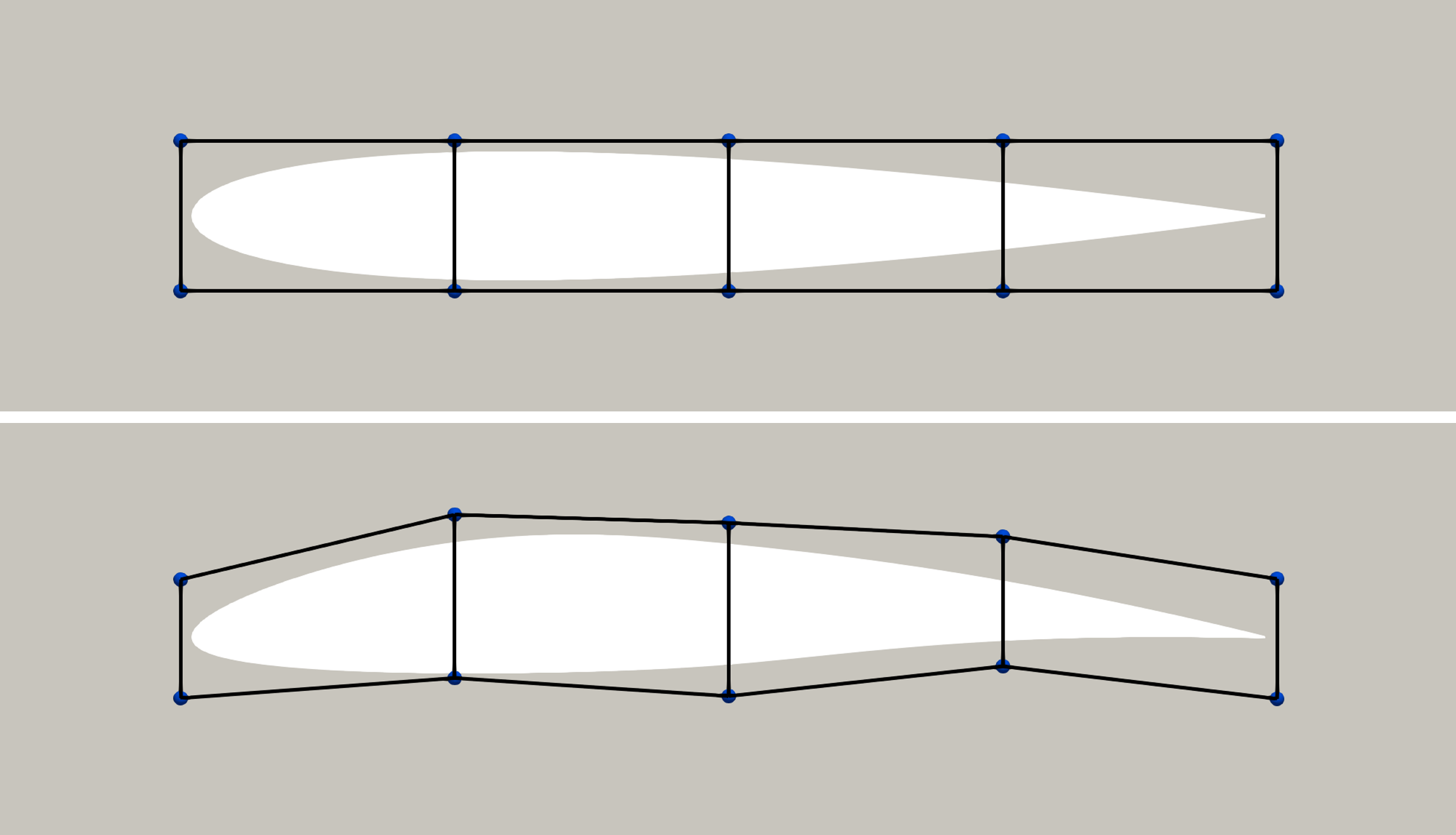}
    \caption{2D example of the FFD method}
    \end{subfigure}
\hspace{0.5em}
    \begin{subfigure}[b]{0.4\textwidth}
    \includegraphics[width=\textwidth]{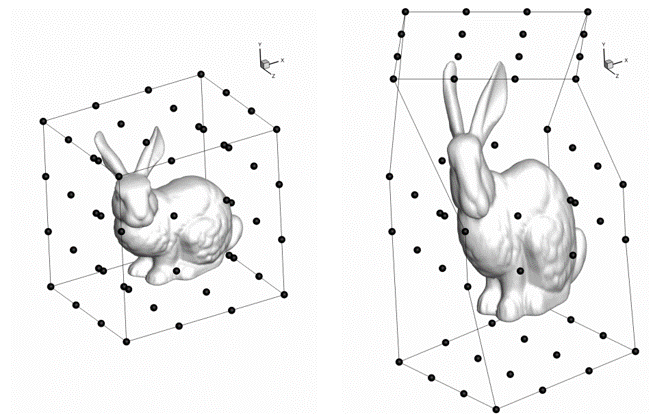}
    \caption{3D example of the FFD method}
    \end{subfigure}
\caption{Examples of geometry deformations with FFD points. (a) a 2D NACA0012 airfoil (from the DAFoam tutorial) and (b) a 3D Stanford bunny (figure taken from~\citet{kenway2010cad}).}    
\label{fig:ffdexamples}
\end{figure}

We need to parametrize the geometry to deform the surface mesh.
In this work, we use the Free-Form Deformation (FFD) method~\cite{sederberg1986free} for the geometry parametrization, implemented in the package pyGeo~\cite{kenway2010cad,reid2021free}. The principle of the FFD method is to enclose the surface mesh nodes in an FFD box with a specified number of control points (also known as FFD points). The FFD points are analytically connected to the enclosed surface nodes using tri-variate $B$-splines.
More details are presented in~\ref{sec:ffd:appendix}. 
Controlling the FFD points enables smooth deformation of the enclosed geometry. 
Figure~\ref{fig:ffdexamples} shows two examples of geometry deformation controlled by the FFD method in two and three dimensions.

\begin{figure}[hbt!]
    \centering
    \includegraphics[width=\textwidth]{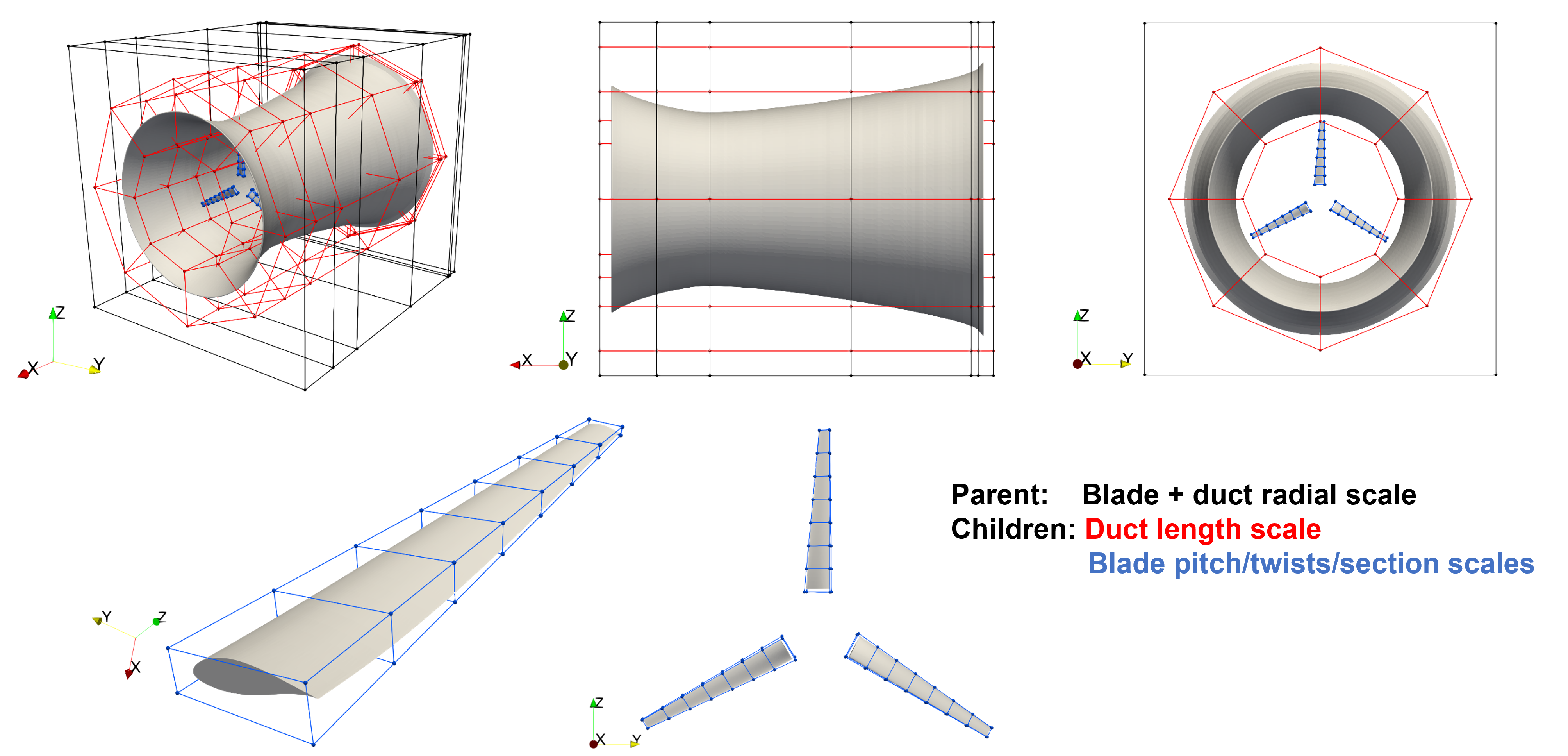}
    \caption{Geometry parametrization of a ducted turbine through the FFD method. Top: different views of the overall FFD setup, where the parental FFD box (black) controls the radial scales of both duct and blades, the child duct FFD box (red) controls the length of the duct, and the child blade (blue) FFD box controls the pitch/twist angles and sectional areas of the blades. Bottom: Closer views of the FFD setup for the blades.}
    \label{fig:geoParam}
\end{figure}

Figure~\ref{fig:geoParam} shows an overview of the FFD setup for our ducted turbine. 
Two levels of FFD boxes are used, with one parent box (black) enclosing all duct and blade geometries and two children boxes (red and blue) enclosing the duct and blades. 
The 21 design variables in Section~\ref{subsec:optproblem} can now be represented by 21 degrees of freedom (DoF) associated with the FFD points. 
The child FFD box for an individual blade has 32 FFD points placed on 8 sections. 
Twist variables rotate the four FFD points about the reference axis located at the quarter chord line. 
Scale variables scale the cross-section by moving the four control points to expand or contract simultaneously.
The FFD points across different blades are linked to ensure the same deformation for all blades. 

The child FFD box for the duct contains 112 FFD points placed on seven sections, but overall only one variable is defined to control all FFD point to change the duct length $l$. 
When changing the duct length, all duct FFD points move in the axial direction with perturbations proportional to their distances from the throat. 
This movement is to ensure that the throat is consistently located at 26.4\% of the overall duct length. Note that seven sections are not necessary. This choice is mainly for convenience during setup. 

The parent FFD box handles the constraint ~\eqref{eq_optconstraint} on the tip-gap ratio and the condition $D_\text{exit}=\sqrt{(4/\pi)A}=1.536m$. Twenty-eight FFD points are placed on seven sections in the parent box, in which the FFD points for the last three sections are closely packed horizontally and remain stationary throughout the optimization, such that $D_\text{exit}=\sqrt{(4/\pi)A}$ is guaranteed. The 16 FFD points on the first 4 sections from the duct inlet are used to control the duct diameters, i.e., design variables $d_i$. As these FFD points move radially, 
the child FFD boxes (enclosed in the parent FFD box) deform and move the embedded surface geometry accordingly.
Therefore, the constraint~\eqref{eq_optconstraint} is automatically satisfied since the blade expands/contracts proportionally to the duct throat. 
This complex FFD setup for ducted turbines, we believe, is novel in the engineering application of the FFD method.

\subsubsection{Adjoint method for derivative computation}
\label{subsubsec:ajoint}
To compute the derivative of an objective function (in this case, the power coefficient $C_P$) with respect to design variables, the adjoint method~\cite{martins2021engineering} is used.
In order to clearly explain the adjoint method, we first introduce notations as follows: Let $\bm{x}\equiv \{ \{ \theta_i, b_i \}_{i=1}^{8}, \{d_j\}_{j=1}^{4}, l\} \in \mathbb{R}^{N_x}$ with $N_x=21$ as the design variables. Let $\bm{s}\in \mathbb{R}^{M}$ be the state variables in the solution of the RANS-MRF equation. Here $M \sim \mathcal{O}(10^7)$ includes three velocities and pressure at each cell in the computational grid. Our goal is to compute $dC_P/d\bm{x} \in \mathbb{R}^{21}$. If a finite-difference method is used to compute the derivative, one needs at least 22 CFD simulations for derivative computation, even with the lowest-order approximation. This is computationally prohibitive for our application.

For the adjoint method to compute $dC_P/d\bm{x}$, the first step is to write the function $C_P(\bm{x},\bm{s}(\bm{x}))$, and express its total derivative with respect to $\bm{x}$ as
\begin{equation}
    \underbrace{\frac{dC_P}{d\bm{x}}}_{1\times N_x}=\underbrace{\frac{\partial{C_P}}{\partial\bm{x}}}_{1\times N_x}+\underbrace{\frac{\partial{C_P}}{\partial\bm{s}}}_{1\times M}\underbrace{\frac{d\bm{s}}{d\bm{x}}}_{M \times N_x},
\label{Adjoint1}
\end{equation}
where $\partial C_P/\partial \bm{x}$ should be considered as the change of power coefficient $C_P$ as the design variables (i.e. geometry) are varied, with flow solution $\bm{s}$ remaining unchanged. $\partial C_P/\partial \bm{s}$ is the change of $C_P$ as the flow solution $\bm{s}$ changes with a fixed turbine geometry. These partial derivatives are relatively easy to compute, with more details presented in~\ref{sec:adjoint:appendix}.  

The term that is difficult to compute in Eq.~\eqref{Adjoint1} is $d\bm{s}/d\bm{x}$. To compute it, one needs to further involve the RANS-MRF state equations in terms of their discretized residual form $\bm{R}(\bm{x},\bm{s}(\bm{x}))=0$.  Here $\bm{R}(\bm{x},\bm{s}(\bm{x})) \in \mathbb{R}^M$ considering the same number of equations as the number of unknowns in $\bm{s}$. Since $\bm{R}(\bm{x},\bm{s}(\bm{x}))$ should remain zero with a change of $\bm{x}$ (if the flow solution is correctly obtained), we have
\begin{equation}
    \underbrace{\frac{d\bm{R}}{d\bm{x}}}_{M\times N_x}=0\Rightarrow \underbrace{\frac{\partial\bm{R}}{\partial\bm{s}}}_{M\times M}\underbrace{\frac{d\bm{s}}{d\bm{x}}}_{M\times N_x}=-\underbrace{\frac{\partial\bm{R}}{\partial\bm{x}}}_{M\times N_x}.
\label{Adjoint2}
\end{equation}
Direct solution of Eq.~\eqref{Adjoint2} gives
\begin{equation}
    \frac{d\bm{s}}{d\bm{x}} = -\underbrace{\left[{\frac{\partial\bm{R}}{\partial\bm{s}}}\right]^{-1}}_{M\times M}\underbrace{\frac{\partial\bm{R}}{\partial\bm{x}}}_{M\times N_x}.
\label{Adjoint3}
\end{equation}
It is worthwhile to discuss the computational cost associated with Eq.~\eqref{Adjoint3} at this point. The matrix multiplication in Eq.~\eqref{Adjoint3} leads to a computational complexity of $\mathcal{O}(M^2 N_x)$ that is very expensive since $M\sim \mathcal{O}(10^7)$ and $N_x$ is also large. This has to be added by the cost to invert a $M\times M$ matrix, which is, in general, more expensive. Even if one uses some iterative solver for linear systems to solve Eq.~\eqref{Adjoint2}, the procedure needs to be repeated for $N_x$ times since $d\bm{s}/d\bm{x}$ (as well as the RHS) has $N_x$ columns. The computation is, therefore, also very expensive.

On the other hand, the computational cost can be significantly reduced by simply substituting Eq.~\eqref{Adjoint3} to Eq.~\eqref{Adjoint1} and considering a re-grouping of the multiplications:
 \begin{equation}
    \underbrace{\frac{dC_P}{d\bm{x}}}_{1 \times N_x}=\underbrace{\frac{\partial{C_P}}{\partial\bm{x}}}_{1 \times N_x}-\left(\underbrace{\frac{\partial{C_P}}{\partial\bm{s}}}_{1 \times M}\underbrace{\left[{\frac{\partial\bm{R}}{\partial\bm{s}}}\right]^{-1}}_{M \times M}\right)\underbrace{\frac{\partial\bm{R}}{\partial\bm{x}}}_{M \times N_x}.
\label{Adjoint4}
\end{equation}
Instead of computing Eq.~\eqref{Adjoint3}, we first compute the multiplication grouped in the parenthesis in Eq.~\eqref{Adjoint4}. This computation can be done by solving the so-called adjoint equation (the adjoint is equivalent to the transpose of a real matrix in our case) 

\begin{equation}
    \underbrace{\left[{\frac{\partial\bm{R}}{\partial\bm{s}}}\right]^{T}}_{M\times M}\underbrace{\bm{\psi}}_{M\times 1}=\underbrace{\left[\frac{\partial{C_P}}{\partial\bm{s}}\right]^{T}}_{M\times 1}
\label{Adjoint5}
\end{equation}
whose solution transpose provides 

\begin{equation}
    \bm{\psi}^{T}=\frac{\partial{C_P}}{\partial\bm{s}}\left[{\frac{\partial\bm{R}}{\partial\bm{s}}}\right]^{-1}
\label{Adjoint6}
\end{equation}
as the parenthesis term in Eq.~\eqref{Adjoint4}.

The solution of Eq.~\eqref{Adjoint5} involves solving a linear system only once, instead of $N_x$ times as needed for Eq.~\eqref{Adjoint2}, and hence is much less expensive (also compared to the direct computation of Eq.~\eqref{Adjoint3}). The computational cost to solve Eq.~\eqref{Adjoint5} is generally similar to the RANS-MRF computation. Therefore, in each iteration of the optimization, the computational cost is in the same order as one RANS-MRF solution. The only remaining component is the calculation of partial derivatives $\partial \bm{R}/\partial \bm{s}$ in Eq.~\eqref{Adjoint5}, which can be found in~\ref{sec:adjoint:appendix} with other derivatives mentioned above.

\subsubsection{Volume mesh deformation}
\label{subsubsec:meshdeform}

When marching to the next design point, the turbine geometry is deformed. The entire computational volume mesh is deformed accordingly. 

\begin{figure}[hbt!]
\centering
    \begin{subfigure}[b]{0.4\textwidth}
    \includegraphics[width=\textwidth]{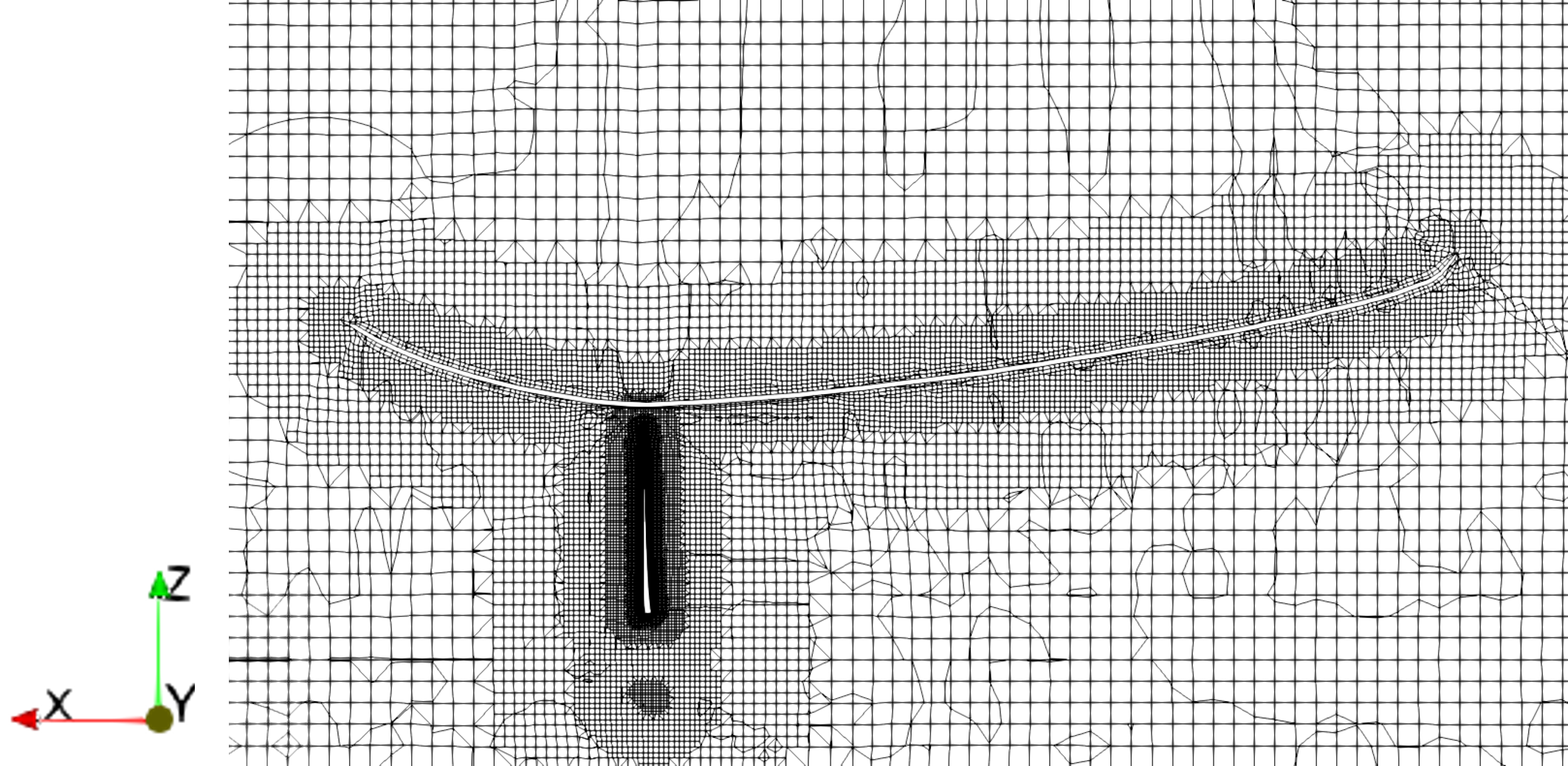}
    \caption{Before deformation}
    \end{subfigure}
\hspace{0.5em}
    \begin{subfigure}[b]{0.4\textwidth}
    \includegraphics[width=\textwidth]{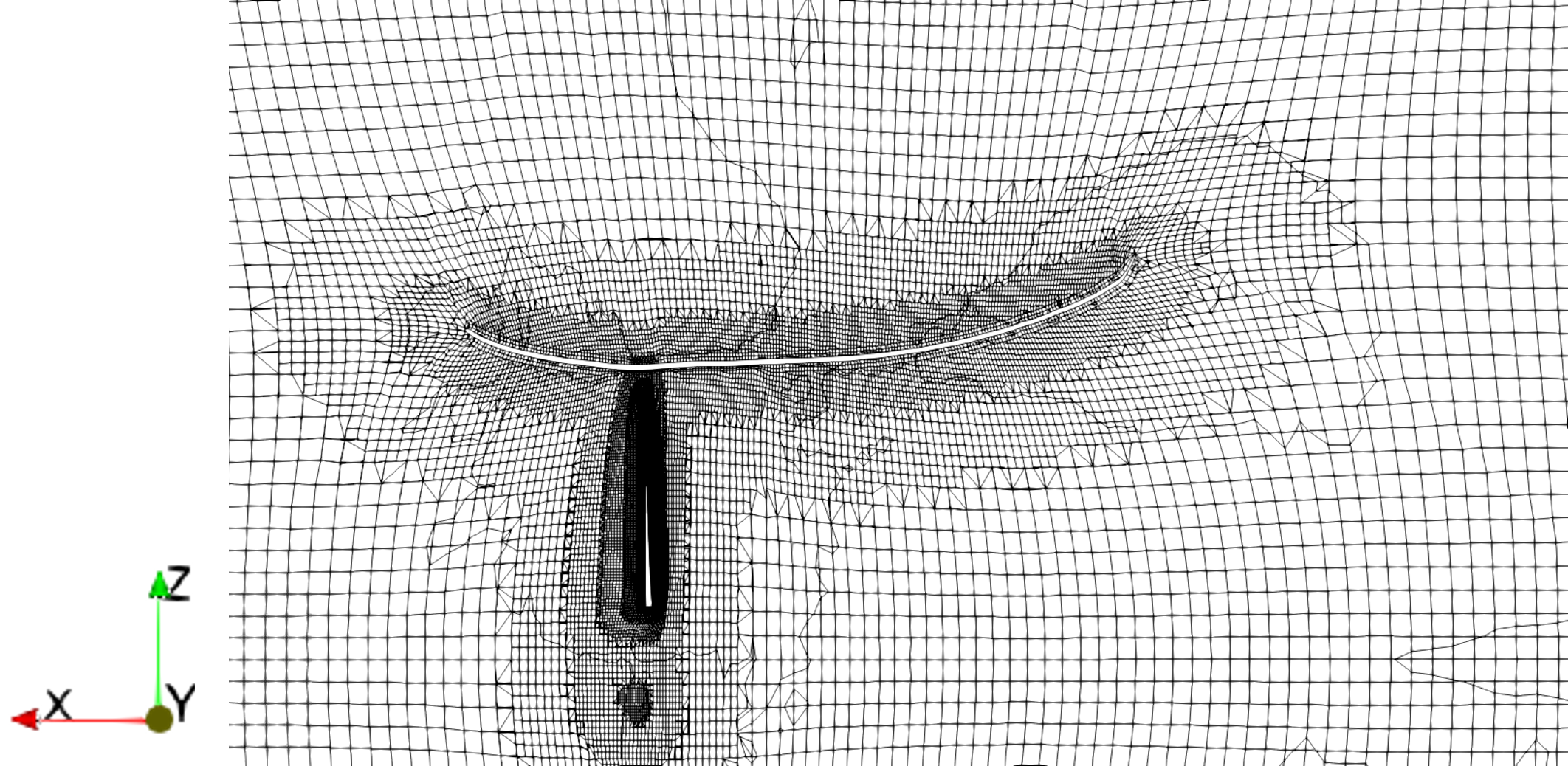}
    \caption{After deformation}
    \end{subfigure}
\caption{Computational mesh deformation during the geometry deformation of a ducted turbine.}
\label{fig:deform}
\end{figure}

The volume mesh deformation is computed based on the analytic inverse-distance weighting method~\cite{luke2012fast}, implemented in the IDWarp package~\cite{Secco2021}. Given a 2D surface, for example a blade surface, with $N$ surface mesh nodes, the geometry deformation leads to the movement of each node. We assign two quantities $(M_i, b_i)$ for each node with $i=1,2,...,N$, where $b_i$ is the translation distance of the node and $M_i$ is the rotation matrix such that $n_i^{new}=M_i n_i^{old}$ with $n_i^{new}$ and $n_i^{old}$ the normal vectors at the node. In particular, both $n_i$'s  are computed by a weighted average of the normal vectors for all surrounding cell faces of the node. After $(M_i, b_i)$ are obtained for $i=1,2,...,N$, we can compute the deformation of any volume mesh by summing the contribution from each surface node, i.e., $\Delta \bm{r}=\sum_{i=1}^N w_i (M_i \bm{r} + b_i - \bm{r})$ with $\bm{r}$ any volume node and $\Delta \bm{r}$ its movement. The weighting factor $w_i$ has the empirical form~\cite{luke2012fast,Secco2021} that grows in a polynomial form with the inverse distance between the volume and surface nodes. Figure~\ref{fig:deform} shows the deformation of the computational mesh during a ducted turbine optimization as an example.

\section{Results and Discussion}
\label{sec:resultdiscussion}
In this section, we present the results of optimization~\eqref{eq:optproblem}, followed by re-evaluation using the higher-fidelity URANS solver, as well as discussions on the optimized geometry and flow mechanism. Before showing the optimization results, we present two additional studies conducted in our work.

\begin{figure}[hbt!]
    \centering
    \includegraphics[width=0.7\textwidth]{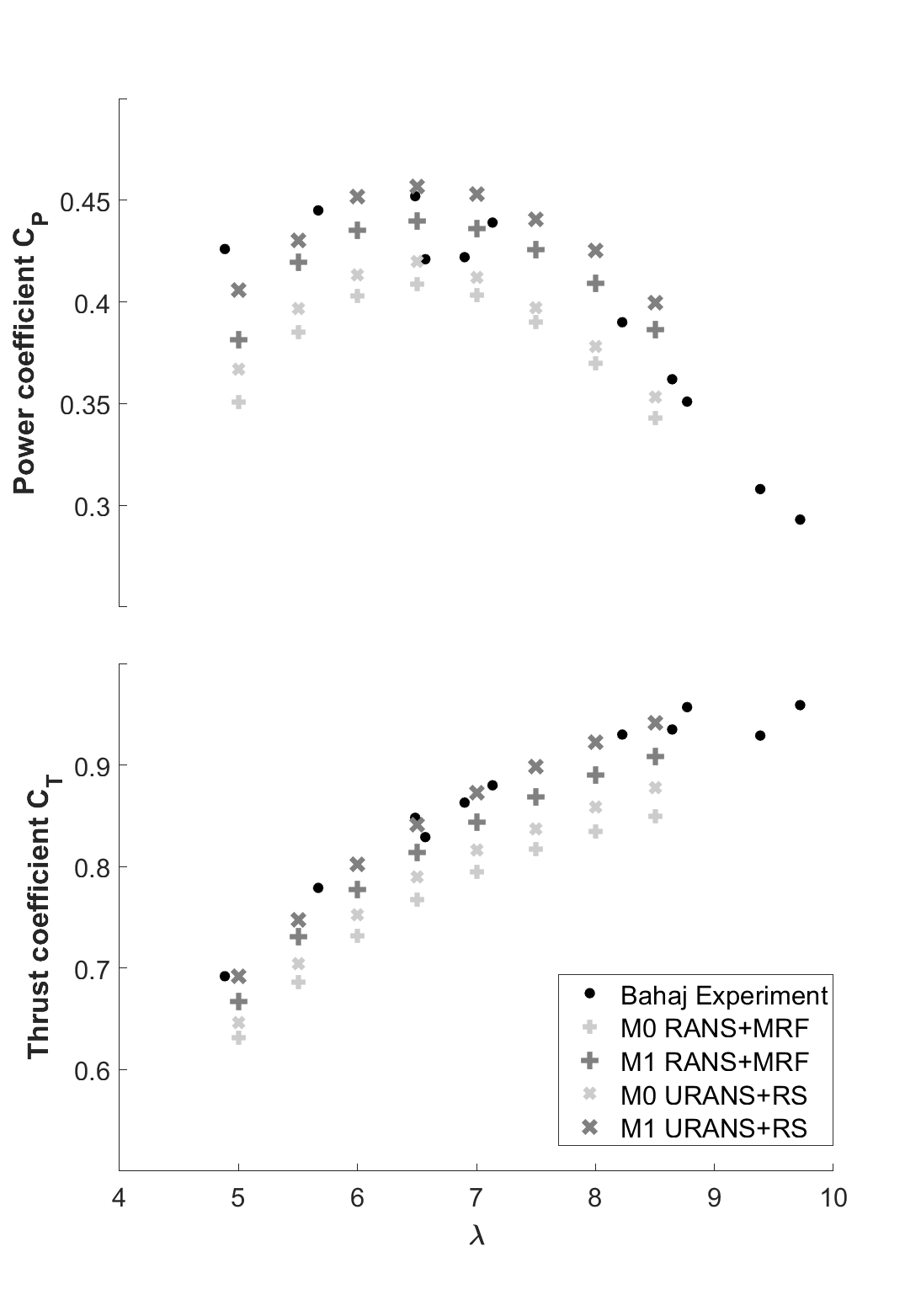}
    \caption{$C_P$ as a function of $\lambda$ for the unducted Bahaj turbine, from the experiment ($\bullet$), the URANS-RS ($\times$), and the RANS-MRF (+).}
    \label{fig:BahajExpvsCFD}
\end{figure}

The first is the validation of the RANS-MRF and the URANS-RS solvers with experimental data. Since we are not aware of systematic measurements of the performance of ducted hydrokinetic turbines, we use the experimental results of the unducted Bahaj turbine for validation. Figure~\ref{fig:BahajExpvsCFD} shows the power coefficient $C_P$ and thrust coefficient $C_T$ (the axial force on the turbine normalized by the momentum of the inflow) for the Bahaj turbine at a range of $\lambda$, obtained from the RANS-MRF and the URANS-RS, in comparison with the experimental results~\cite{bahaj2007power}, all at experimental Reynolds number $Re=1\times10^{6}$. The RANS-MRF solver is run with a second-order numerical scheme for the convection term in the RANS equations, which will be changed in the optimization process as described later in detail. From Figure~\ref{fig:BahajExpvsCFD}, it is clear that for the unducted Bahaj turbine, both solvers predict similar results mostly consistent with the experimental data. The URANS solver seems more accurate in evaluating $C_T$ for all $\lambda$ and $C_P$ at lower $\lambda$ (e.g., the value on which our optimization is based).


The second is a grid-search study with 5 design parameters using the RANS-MRF and coarse-grid (even coarser than M0) flow solvers that we conducted before the gradient-based optimization. 
This search, as detailed in~\ref{sec:discreteDesign:appendix}, does \emph{not} provide a successful design of the ducted hydrokinetic turbine with improved efficiency (compared to the unducted Bahaj model). This failure is very likely due to the low-dimensional parameter space, which is insufficient to explore effective designs. It also implies that designs with more parameters using a gradient-based method, as we present below, are necessary for designing complex geometries such as ducted turbines.  

\subsection{Optimization and Re-evaluation}
\label{sec:result_flowrate}
We solve the optimization problem~\eqref{eq:optproblem} using methods described in Section~\ref{subsec:optMethod}. 
In the RANS-MRF solver, a first-order numerical scheme is used for computing the convective term (i.e., to construct the flux in cell faces). We note that the first-order scheme is more dissipative than the normally-used second-order scheme, but the former is critical to obtain convergent flow solutions for many duct designs, especially those associated with flow separation. Specifically, upon extensive tests, we find that the second-order scheme shows fluctuating flow solutions in many cases, which in turn affects the accuracy of the adjoint method, preventing an accurate gradient computation. On the other hand, while the first-order scheme may provide less accurate solutions for cases with separated flow, the obtained $C_P$ is usually low for these cases and the optimization leads to designs with no flow separation and with improved efficiency. In each RANS-MRF simulation, we consider the solution converged as the residuals stop dropping, which in general occurs when residuals of momentum equation reach $\mathcal{O}(10^{-4}\sim 10^{-5})$.

\begin{figure}[hbt!]
    \centering
    \includegraphics[width=0.8\textwidth]{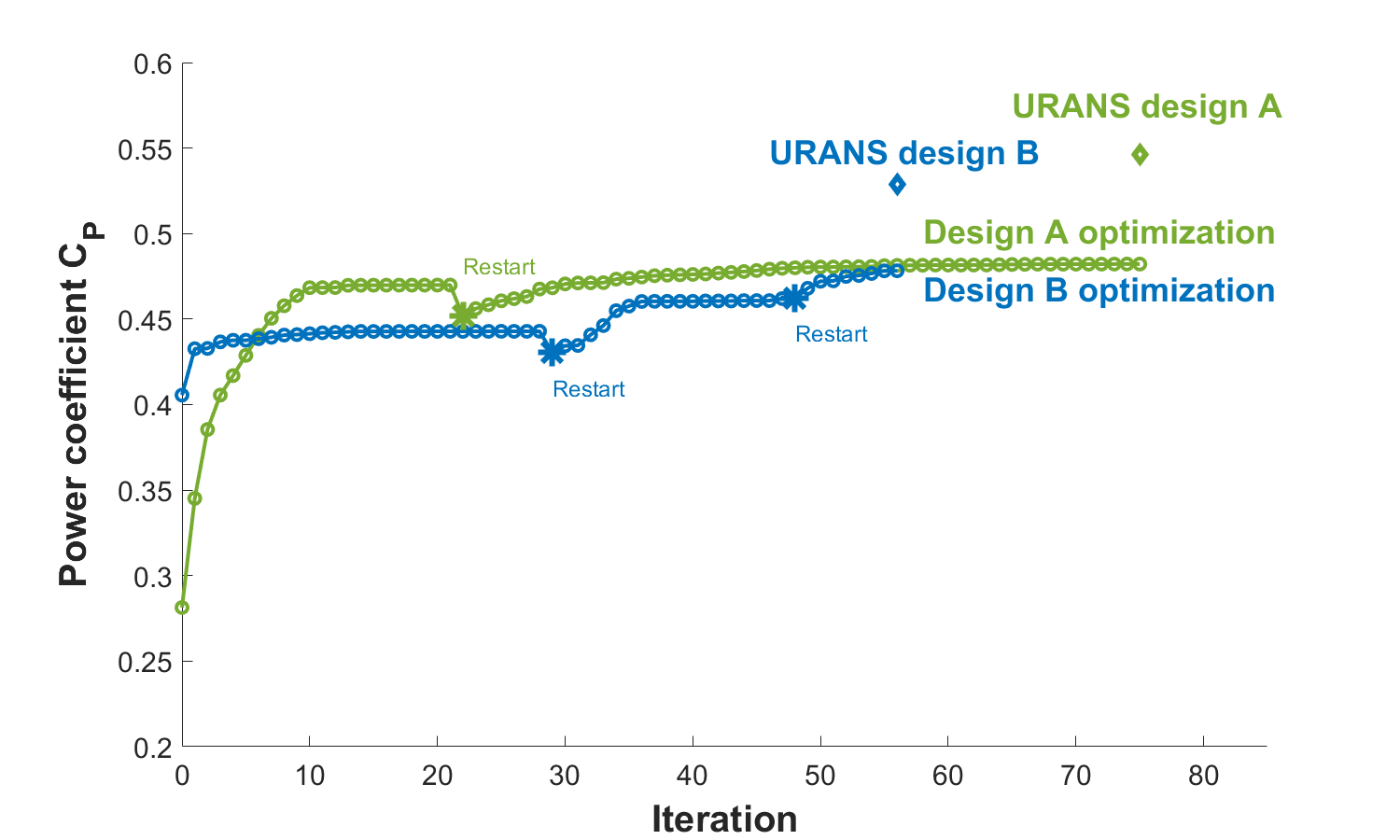}
    \caption{History of $C_P$ for optimizations A (green) and B (blue). The URANS evaluation of $C_P$ for optimized designs A and B is shown by diamond symbols. The restarting points are denoted as asterisks($\ast$).}
    \label{fig:iterCP}
\end{figure}

Figure~\ref{fig:iterCP} shows the change of $C_P$ as the optimization progresses, starting from both baseline designs A and B (hereafter optimizations A and B). We first notice that the two starting points yield $C_P=25\%$ and $41\%$, respectively, evaluated by the RANS-MRF. 
Both values are lower than the counterparts ($28\%$ and $45\%$) reported earlier from the URANS-RS. We see that optimization of both baseline designs leads to a fast increase of $C_P$ at the beginning until both $C_P$ values (almost) plateau. The small bumps on both curves in Figure~\ref{fig:iterCP} correspond to the restarting/re-meshing procedure. This remeshing step is necessary because geometry deformations that are too large lead to mesh quality degradation and hence deteriorate the quality of the flow and adjoint solutions. A manual restarting/re-meshing procedure improves the optimization behavior, leading to additional increases of $C_P$. Both optimizations A and B are stopped when the $C_P$ value plateaus even with further restarting/re-meshing. In practice, we find that this convergent situation corresponds to the SNOPT optimality metric~\cite{gill2005snopt} of approximately $10^{-1.7}$, which is consistent with cases in an unducted wind turbine optimization~\cite{dhert2017aerodynamic}. 
For such optimality conditions, although $\nabla C_P$ does not completely vanish, the benefit of further optimizing the turbine is compromised by the mesh deformation so that some practical optimal points are reached. The two optimized designs yield similar values of $C_P$, namely $0.4822$ from A and $0.4782$ from B achieved respectively at $\lambda=6.39$ and $\lambda=6.18$ (since the blade radius $R$ is optimized, which affects $\lambda$). 

Since the RANS-MRF is the low-fidelity solver (due to issues mentioned in Section~\ref{subsec:mrf} and the first-order convection scheme), we re-evaluate the optimized designs A and B using the high-fidelity URANS-RS. The obtained values from the URANS-RS on the M0 grid are added to Figure~\ref{fig:iterCP}, which yield $54.6\%$ and $52.9\%$ for the optimized designs A and B. Here, the difference between two solvers for ducted turbines (due to the complexity of the flow) is remarkably larger than that for unducted turbines as shown in Figure~\ref{fig:BahajExpvsCFD}. 

A grid sensitivity study is also conducted, which evaluates $C_P$ for the optimized designs A and B using the URANS-RS on meshes M0, M1, and M2 with increasing resolution. In these URANS solutions, an adaptive time step is used (with an average time step of $1\times10^{-5}$ seconds, i.e., about $0.01^{\circ}$ turbine rotation for 1 time step) with total simulation times of 15 seconds to reach the quasi-steady state. With the NSF Stampede2 cluster, the simulations using the M2 grid take about 240 hours on 576 CPUs. The results are shown in Table~\ref{table:URANSresult}, which indicates that $C_P$ values vary only by $\mathcal{O}(1\%)$ (in terms of the absolute value), i.e., they are not very sensitive to the large range of variation of grid resolutions. Based on results from M2, the two final designs yield similar $C_P \approx 54\%$ that is much higher than $46\%$ of the unducted Bahaj model (or standard unducted turbines).

\begin{table}[h!]
\centering
\begin{adjustbox}{width=\columnwidth,center}
\begin{tabular}{lcccc}
\multicolumn{5}{c}{\textbf{Design A}}\\ 
\hline
    & \textbf{Num of cells} & \textbf{$y^{+}$ Blade} & \textbf{$y^{+}$ Duct} & \textbf{$C_P$} \\
    \hline
    \textbf{M0} & 2,741,276 & 47.75 & 200.2 &0.5462 \\
    \textbf{M1} & 4,697,325 & 40.82 & 191.1 &0.5511 \\
    \textbf{M2} & 7,280,862 & 33.30 & 147.7 &0.5381 \\
\hline\hline
\end{tabular}
\quad
\begin{tabular}{lcccc}
\multicolumn{5}{c}{\textbf{Design B}}\\ 
\hline
    & \textbf{Num of cells} & \textbf{$y^{+}$ Blade} & \textbf{$y^{+}$ Duct} & \textbf{$C_P$} \\
    \hline
    \textbf{M0} & 3,066,956 & 49.32 & 221.1 &0.5287 \\
    \textbf{M1} & 5,358,847 & 40.81 & 173.5 &0.5335 \\
    \textbf{M2} & 8,453,165 & 34.33 & 155.4 &0.5337 \\
\hline\hline
\end{tabular}
\end{adjustbox}
\caption{URANS-RS re-evaluation of the optimized design A (left) and B (right) with grid sensitivity studies, including cell numbers, $C_P$, as well as $y^+$ values on both duct and blades. These $y^+$ values are in the applicable range of the automatic wall treatment (Appendix A).}
\label{table:URANSresult}
\end{table}

\begin{figure}[hbt!]
    \centering
    \includegraphics[width=0.8\textwidth]{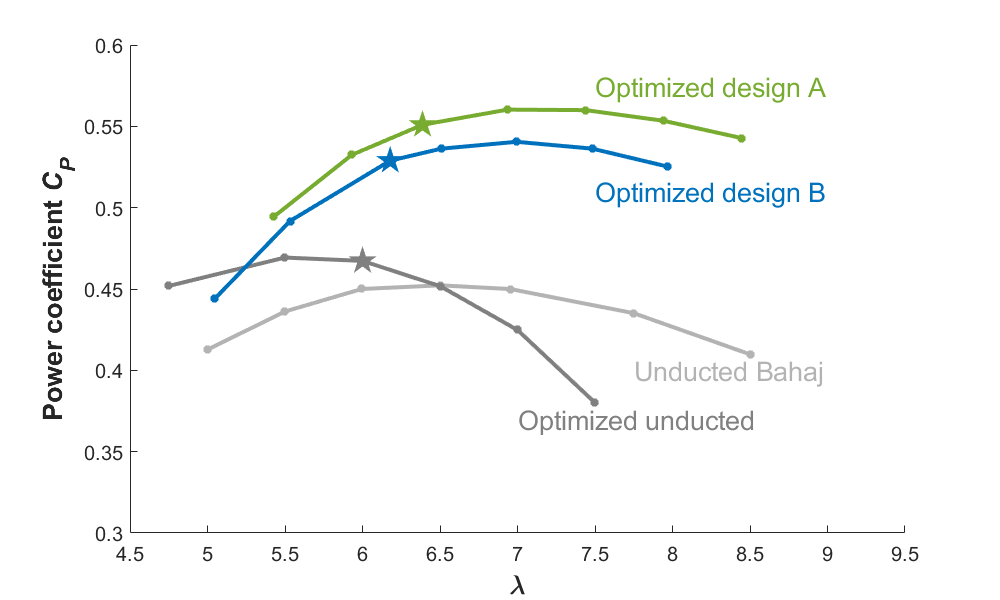}
    \caption{$C_P$ as a function of $\lambda$ for unducted Bahaj turbine (gray), optimized unducted turbine (dark gray), optimized ducted turbine A (green), and optimized ducted turbine B (blue). The design points (at a given rotating speed) are marked by stars.}
    \label{fig:TSRvsCP}
\end{figure}

We further evaluate $C_P$ of the two optimized designs at a range of $\lambda$ using the URANS-RS with the M0 grid. The results are shown in Figure~\ref{fig:TSRvsCP} together with $C_P$ of the unducted Bahaj model, as well as the optimized unducted turbine using the same setup. We note that the unducted turbine is optimized for fixed $\Omega = 21\ rad/s$, corresponding to $\lambda=6$. We see that the optimized ducted turbine designs not only work well for the designed value of $\Omega$ but also perform with high efficiency for a large range of $\lambda$. Moreover, the maximum $C_P$ for each design is in fact not achieved at the designed $\lambda$ (marked by stars in the figure) but at some larger value of $\lambda$. Considering Figure~\ref{fig:TSRvsCP}, the maximum $C_P$ for the two designs are $56\%$ and $54\%$ achieved at $\lambda=6.94$ and $6.99$, respectively.

\begin{figure}[hbt!]
    \centering
    \includegraphics[width=0.8\textwidth]{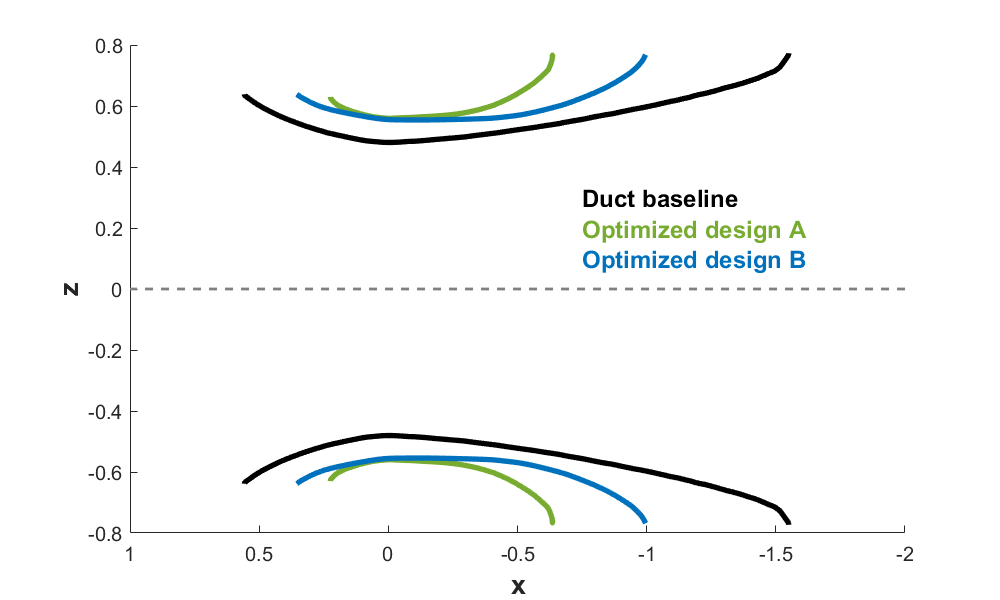}
    \caption{Duct designs, including the baseline (black), optimized design A (green), and optimized design B (blue).}
    \label{fig:opt_duct_geo}
\end{figure}

We finally examine the geometries of the optimized designs A and B. Figure~\ref{fig:opt_duct_geo} shows the optimized duct shapes of the two designs, laid on top of the baseline duct design~\cite{Knight2018}. Overall, we see that the optimized designs have shorter duct lengths and enlarged throat area.
The optimized design A has a duct with length of $0.861m$ ($59.1\%$ reduction compared to the baseline ducted turbine) and a throat radius of $0.559m$ ($16.5\%$ increase).
The optimized design B has a duct with length of $1.350m$ ($35.9\%$ reduction compared to the baseline ducted turbine) and a throat radius of $0.554m$ ($15.5\%$ increase).

\begin{figure}[hbt!]
    \centering
    \includegraphics[width=0.9\textwidth]{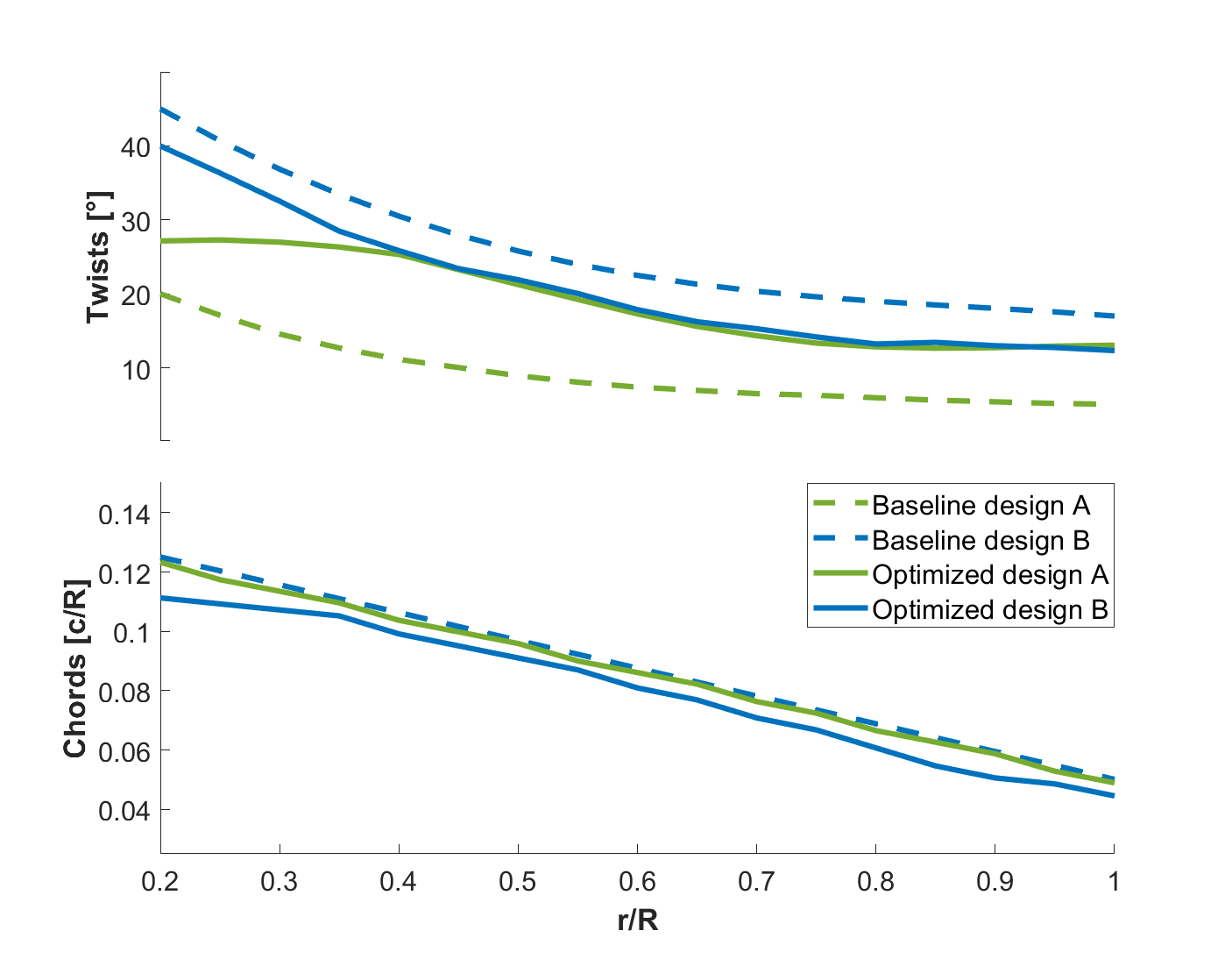}
    \caption{Distributions of twists (top) and chord lengths (bottom) for baseline designs A (dashed green) and B (dashed blue), optimized designs A (solid green) and B (solid blue).}
    \label{fig:opt_twistchord}
\end{figure}

The twist and chord length profiles of the optimized/baseline designs are shown in Figure~\ref{fig:opt_twistchord}. We see that the chord lengths of the optimized designs do not vary much from the baseline, but significant changes occur in the twist profile through the optimization. Moreover, starting from drastically different twist profiles in baseline designs A and B, the two optimized designs converge to very similar twist distributions, especially for $r/R>0.35$, where most torque is generated. 
Different geometries of the optimized designs A and B may indicate that they locate on two local optima in the design parameter space. Given the comparable performance despite the different duct lengths, we conclude that the blade twists and duct throat areas are the driving design parameters.

\subsection{Analysis of Flow Mechanism}
\begin{figure}[hbt!]
    \centering
    \includegraphics[width=0.8\textwidth]{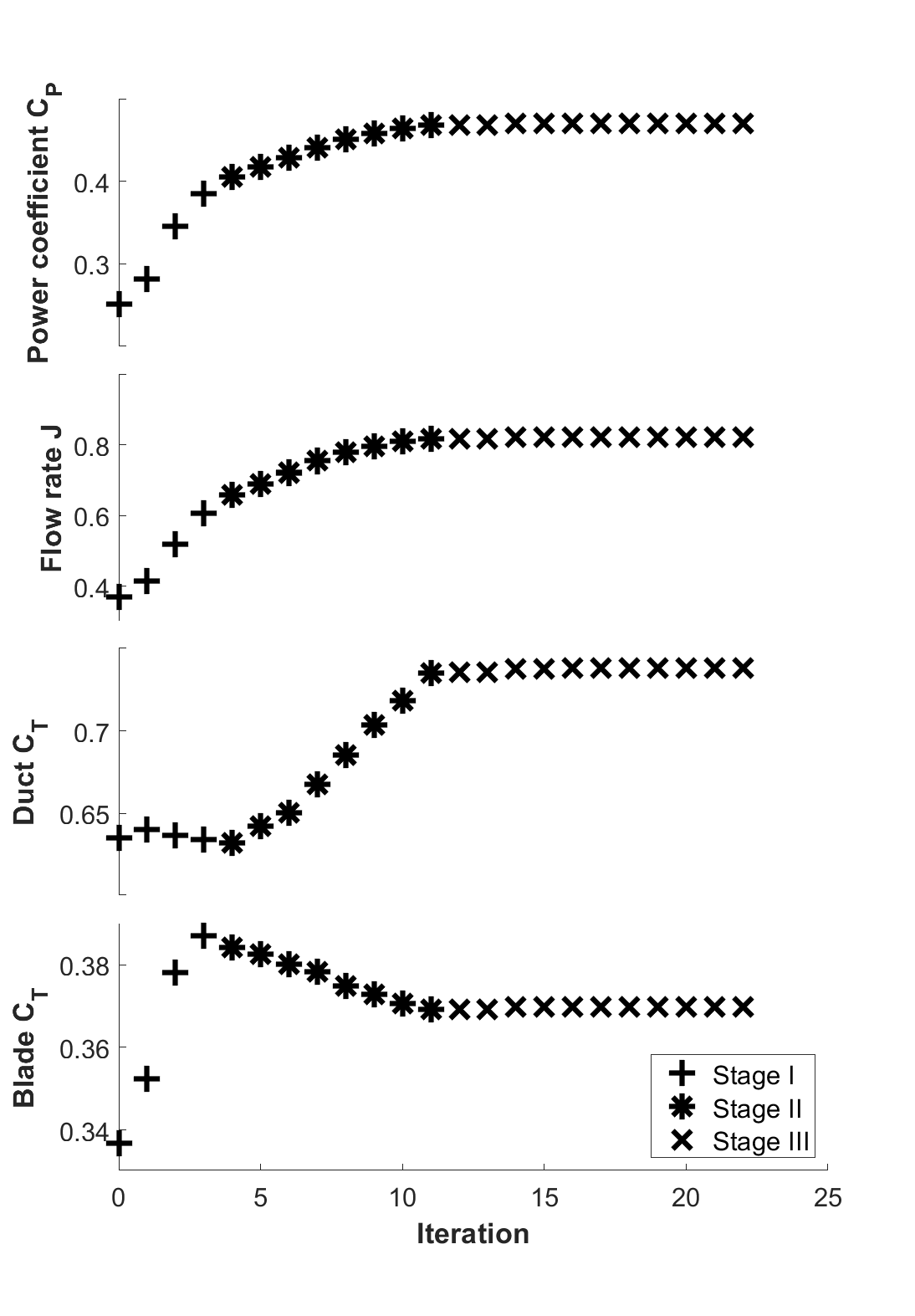}
    \caption{Variations of $C_P$ (first row), flow rate $J$ (second row), duct $C_T$ (third row), and blade $C_T$ (fourth row) in the first 22 iterations of the optimization. Three stages of the fast growth of $C_P$, slow growth of $C_P$, and plateau of $C_P$ are marked by plus sign(+), asterisk($\ast$), and cross($\times$).}
    \label{fig:values_vs_iter}
\end{figure}

In this section, we analyze the flow fields of unducted turbines and baseline/optimized ducted turbines in order to understand the major flow mechanism leading to the improvement of performance. The optimized design A is used as an illustration. We first plot in Figure~\ref{fig:values_vs_iter} variations of some relevant performance metrics together with the variation of $C_P$ (first row) in the first 22 iterations of the optimization process (before the first restarting/re-meshing). These metrics include the flow rate $J$ passing the turbine blades and the thrust coefficient $C_T=T/(0.5\rho U_\infty^2 A)$ on the duct and blades, shown respectively in the second, third, and fourth rows of Figure~\ref{fig:values_vs_iter}. 

We also divide the 22 iterations of the optimization process into three stages I, II, and III, respectively: stages with the fast growth of $C_P$ (marked by `+' in the figure), slow growth of $C_P$ ($\ast$), and plateau of $C_P$ ($\times$). 

In stage I, while the duct $C_T$ remains almost unchanged, the blade $C_T$ increases rapidly. This is the most favorable situation to improve $C_P$ since clearly more and more loading from the total is distributed on the blades. In stage II, this favorable variation of $C_T$ cannot be maintained (i.e., its potential has been exhausted in stage I), and the opposite trend is observed with increased duct $C_T$ and decreased blade $C_T$. The further (slow) increase of $C_P$ in stage II, therefore, must be associated with a different mechanism that is perhaps the more effective transition from blade loading to rotational motion (or torque). 
Both duct and blade $C_T$ become unchanged in stage III as $C_P$ plateaus. The overall increase of $C_T$ in the whole process is $14\%$ and the $C_P$ increase is $87\%$.
This is another favorable feature of the current optimization since it would be much more demanding for supporting structures with a much higher $C_T$. Finally, the flow rate $J$ is highly correlated with $C_P$ in the whole optimization process and increases constantly until its simultaneous plateau with $C_P$.

\begin{figure}[hbt!]
    \centering
    \begin{subfigure}[b]{0.8\textwidth}
    \includegraphics[width=\textwidth]{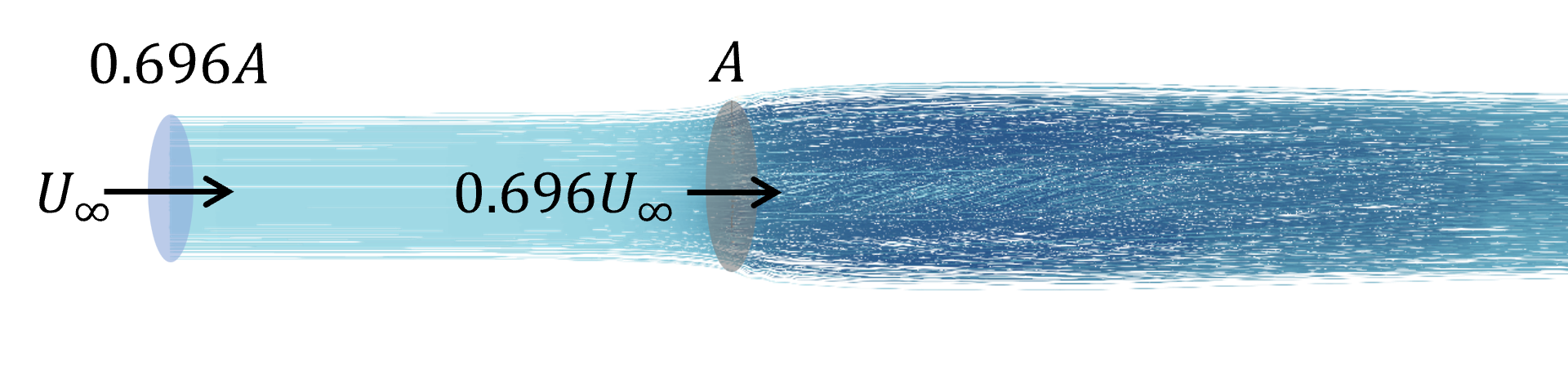}
    \caption{Optimized unducted turbine}
    \end{subfigure}
    
    \begin{subfigure}[b]{0.8\textwidth}
    \includegraphics[width=\textwidth]{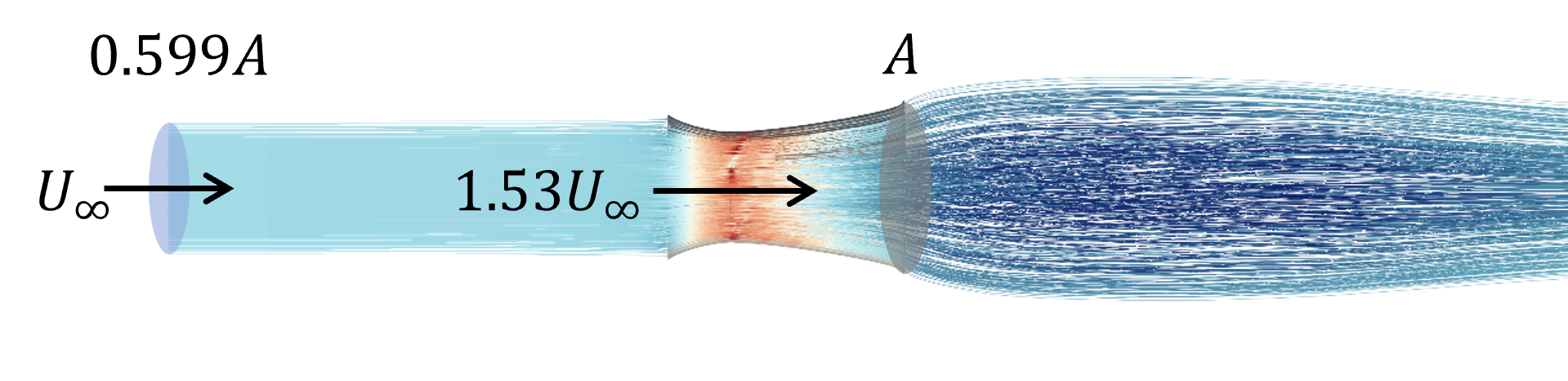}
    \caption{Baseline design A}
    \end{subfigure}
    
    \begin{subfigure}[b]{0.8\textwidth}
    \includegraphics[width=\textwidth]{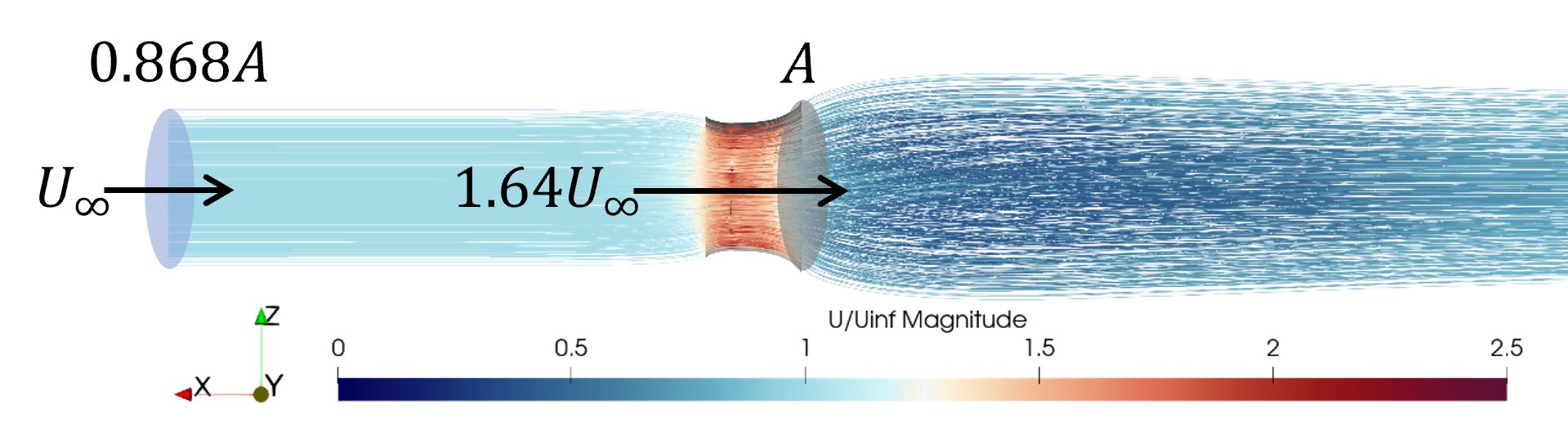}
    \caption{Optimized design A}
    \end{subfigure}
    
\caption{Streamlines in a flow tube passing the blades of (a) the optimized unducted turbine, (b) baseline design A, and (c) optimized design A.}
\label{fig:streamline}
\end{figure}

The above analysis motivates us to study further the flow rate metric, which remains consistent with the trend of $C_P$ in the optimization process. In Figure~\ref{fig:streamline}, we show flow visualizations for the optimized unducted turbine, baseline turbine A, and optimized turbine A, obtained in the quasi-steady solution from the URANS solver. To facilitate a fair comparison, we only show the streamlines in the flow tube that passes the turbine blades. Since the inflow velocity is fixed at $1.4m/s$ for all cases, the flow rate in the tube is proportional to the area of the tube at the inlet.

From Figure~\ref{fig:streamline}, it is clear that the optimized design corresponds to the case with the largest flow tube inlet area. Physically, this indicates that a well-designed duct draws a larger volume of water (compared to unducted and baseline turbines) into the throat, which is accompanied by a higher flow rate ($2.25m^3/s$ compared to $1.80m^3/s$ and $1.55m^3/s$ for the other two cases) across the blades. This metric of flow rate is the most effective indicator of the ducted turbine performance, instead of the flow speed at the throat. As an example, the baseline design A is associated with an accelerated flow speed at the throat but not improved efficiency. This analysis explains the enhanced performance of the optimized ducted turbine through the improved flow conditioning provided by the duct, confirming the long-existing hypothesis in the field of hydrokinetic turbines. 

\section{Conclusions}
\label{sec:conclusion}
In this paper, we conduct gradient-based design optimization of ducted hydrokinetic turbines using CFD and the adjoint method. Two baseline designs with drastically different performances are chosen as starting points of the optimization. The resultant designs for both cases yield similar performance with $C_P\approx 54\%$ when evaluated by the high-fidelity URANS solver.
Both designs capture similar critical geometrical features in terms of the duct throat area and blade twist profile. This value of $C_P$ is 8\% higher than standard unducted turbines, including the Bahaj model. We further demonstrate that the optimized designs not only achieve high $C_P$ at the design rotational speed, but also yield high performance over a wide range of rotating speeds and thus $\lambda$. Finally, we study the flow mechanism associated with performance improvement and show that $C_P$ among different designs is correlated to the flow rates passing the turbine blades. The optimized design corresponds to the case with a maximum flow rate due to the suction of a well-designed duct.

The current work demonstrates the great potential of gradient-based optimization with the adjoint method in designing geometrically complex renewable energy devices such as ducted turbines. Nevertheless, the current optimized design needs further modifications for model tests and real-world applications. Two major issues of the current design are: (1) it has a thin-wall-shaped duct that is difficult to manufacture; (2) it does not include a hub that is necessary for installation. Both issues result from the limitation of the FFD method for geometry parametrization that one needs to improve for better designs. We are now working on the Engineering Sketch Pad (ESP)~\cite{haimes2013engineering} for geometry parametrization, which can, in principle, overcome the above two issues. With the ESP method replacing the FFD method, we expect our next-round design to result in some geometry that is ready for a model test in the towing tank at the University of Michigan.

\section{Acknowledgements}
\label{sec:acknowledge}
This work is a part of the Re-configurable Array of High-Efficiency Ducted Turbines for Hydrokinetic Energy Harvesting (RAFT) project, supported by the United States Department of Energy (DOE)-ARPA-E under SHARKS program award No. DE-AR0001438 (Program Director Dr. Mario Garcia-Sanz). We thank the ARPA-E staff, especially SHARKS program manager Dr. Mario Garcia-Sanz, for the financial support and helpful discussions and for challenging us to improve the quality of the work. We also thank the full RAFT team, especially Dr. M. Reza Amini, Dr. Kartik Praful Naik, and Mr. Boxi Jiang for many discussions (weekly within the University of Michigan and bi-weekly for the whole team). Help from Dr. Ping He on DAFoam is also greatly appreciated.

This work utilized the Stampede2 high performance computing (HPC) system at the University of Texas at Austin as well as the Anvil HPC system at Purdue University through allocation TG-MCH220016. This computing allocation is supported by the Extreme Science and Engineering Discovery Environment (XSEDE), which is supported by National Science Foundation grant number \#1548562 and also by the Advanced Cyberinfrastructure Coordination Ecosystem: Services \& Support (ACCESS) program, which is supported by National Science Foundation grants \#2138259, \#2138286, \#2138307, \#2137603, and \#2138296.

\appendix

\section{$k-\omega$ SST turbulence model and automatic wall treatment}
\label{sec:kwsst:appendix}

The approximation of the Reynolds stress term in Eq.~\eqref{eq_RANS} can be obtained by the Boussinesq hypothesis, which assumes that the Reynolds stress is related to an eddy-viscosity $\nu_t$ and the mean velocity gradients, namely (for incompressible flow)
\begin{equation}
        -\overline{u_i'u_j'} = \nu_t\left(\frac{\partial \overline{U_i}}{\partial x_j}+\frac{\partial \overline{U_j}}{\partial x_i} \right) - \frac{2}{3}k\delta_{ij}
\label{eq_Boussinesq}
\end{equation}
where $k=\frac{1}{2}\overline{u_i'u_i'}$ is the turbulent kinetic energy and $\delta_{ij}$ is the Kronecker delta function. To obtain $k$ and $\nu_t$ in Eq.~\eqref{eq_Boussinesq}, two-equation turbulence models based on $k-\epsilon$ model~\cite{jones1972prediction} or $k-\omega$ model~\cite{wilcox1988reassessment} have been developed. Here, $\epsilon$ is the turbulence dissipation rate, and $\omega = \epsilon/(k C_{\mu})$ with $C_{\mu}=0.09$ is the so-called specific turbulence dissipation rate. These models include transport equations for $k$ and $\epsilon$ (or $\omega$), and $\nu_t$ is obtained as a function of the computed $k$ and $\epsilon$ (or $\omega$) values.

It has been identified by~\citet{wilcox1993comparison} that the $k-\epsilon$ model is robust for regions far from the wall but not accurate when integrating down to the wall (some wall functions are necessary to simulate the viscous sublayer). On the other hand, the $k-\omega$ model offers a more accurate resolution of the viscous sublayer but is less accurate for the far field, e.g., it is over-sensitive to the freestream turbulent conditions~\cite{kok2000resolving, menter2003ten, menter2009review}. On top of both models, the $k-\omega$ based Shear Stress Transport (SST) model~\cite{menter2003ten,menter1994two} is proposed, which connects the near-wall region predicted by the $k-\omega$ model and far-field by the $k-\epsilon$ model using blending functions. In principle, the $k-\omega$ SST model captures the advantage of both models.

In addition to the $k-\omega$ SST model, we apply an automatic wall treatment~\cite{kalitzin2005near} for near-wall simulation. The automatic wall treatment offers a wall $y^+$-insensitive simulation, i.e., great flexibility in the first-cell size at the wall. The principle is to set up different wall boundary conditions for $k$ and $\omega$ depending on the $y^+$ value. It has been verified~\cite{menter2003ten,kalitzin2005near} that the flow prediction in the near-wall region is insensitive and robust for $y^+$ from $\mathcal{O}(0.1)$ to $\mathcal{O}(100)$, and the automatic wall treatment has been practically applied for cases with the first cell up to the range in the logarithmic layer, i.e., $y^+ < 300$.

\section{Derivation of equations in the rotating reference frame}
\label{sec:mrf:appendix}
We start by stating transformations of velocity and acceleration in inertial and rotating reference frames
\begin{linenomath}
\begin{align}
    \bm{U} &= \bm{U}_R+ \bm{\Omega}\times\bm{r} \label{eq:MRF:velocity} \\
    \frac{d\bm{U}}{dt} &= \left[\frac{d\bm{U}}{dt}\right]_R + \frac{d\bm{\Omega}}{dt}\times\bm{r}+2\bm{\Omega}\times\bm{U}_R+\bm{\Omega}\times\bm{\Omega}\times\bm{r},
\label{eq:MRF:acceleration}
\end{align}
\end{linenomath}
where variables with subscript $R$ are evaluated in the rotating reference frame. Here we consider the case $d\bm{\Omega}/dt=0$ that is consistent with our application. Derivations of Eq.~\eqref{eq:MRF:acceleration} can be found in standard textbooks on dynamics, such as~\cite{taylor2005classical}.

Substituting Eq.~\eqref{eq:MRF:velocity} into Eq.~\eqref{eq_NSmass} and making use of the fact that divergence of a curl of a vector is zero, we obtain the continuity equation in rotating reference frame $\nabla\cdot\bm{U}_R = 0$. In order to derive momentum equations in the rotating reference frame, we consider Eq.~\eqref{eq_NSmomentum} with material derivatives replacing the unsteady and convection terms, to which we can then substitute Eq.~\eqref{eq:MRF:acceleration} to obtain
\begin{equation}
    \left[\frac{d\bm{U}}{dt}\right]_R + 2\bm{\Omega}\times\bm{U}_R + \bm{\Omega}\times\bm{\Omega}\times\bm{r} = -\frac{1}{\rho}\nabla p + \nabla\cdot(\nu\nabla\bm{U})
\label{eq_MRF_NSrotating1}
\end{equation}

The viscous term $\nabla\cdot(\nu\nabla\bm{U})$ can be transformed to $\nabla\cdot(\nu\nabla\bm{U}_R)$ using the fact that $\nabla\cdot\left[\nu\nabla\left(\bm{\Omega}\times\bm{r}\right)\right]=0$. We then expand the material derivative in Eq.~\eqref{eq_MRF_NSrotating1} to obtain the momentum equations in the rotating reference frame:
\begin{equation}
    \frac{\partial\bm{U}_R}{dt}+\nabla\cdot(\bm{U}_R\bm{U}_R) = -\frac{1}{\rho}\nabla p + \nabla\cdot(\nu\nabla\bm{U}_R) - 2\bm{\Omega}\times\bm{U}_R - \bm{\Omega}\times\bm{\Omega}\times\bm{r}
\label{eq_MRF_NSrotating}
\end{equation}

While Eq.~\eqref{eq_MRF_NSrotating} can be considered the final equation in a rotating reference frame, it can be simplified further for easier implementation in the finite volume method. For this purpose, we re-arrange the convection term as
\begin{equation} 
\begin{split}
\nabla\cdot (\bm{U}_R\bm{U}_R) & = \nabla\cdot [\bm{U}_R(\bm{U}-\bm{\Omega}\times\bm{r})] \\
 & = \nabla\cdot (\bm{U}_R\bm{U}) - \nabla\cdot\bm{U}_R(\bm{\Omega}\times\bm{r}) - \bm{U}_R\cdot\nabla(\bm{\Omega}\times\bm{r}) \\
 & = \nabla\cdot (\bm{U}_R\bm{U}) - \bm{\Omega}\times\bm{U}_R 
\end{split}
\label{eq_MRF_convection}
\end{equation}
where the following equality is used for the third line
\begin{equation} 
\begin{split}
\bm{U}_R\cdot\nabla(\bm{\Omega}\times\bm{r})
 & = U_{R_l}\frac{\partial}{\partial x_l}\epsilon_{ijk}\Omega_j r_k \\
 & =  U_{R_l} \epsilon_{ijk}\Omega_j \frac{\partial r_k}{\partial x_l}\\
 & =  U_{R_l} \epsilon_{ijk}\Omega_j \delta_{kl}\\
 & =  \epsilon_{ijl}\Omega_j U_{R_l} \\
 & =  \bm{\Omega}\times\bm{U}_R.
\end{split}
\label{eq_MRF_convection2}
\end{equation}
The final equations in the rotating reference frame therefore yield
\begin{equation}
    \begin{cases}
        \nabla\cdot \bm{U} = 0 \\
        \frac{\partial\bm{U}_R}{\partial t} +\nabla\cdot (\bm{U}_R\bm{U}) = -\frac{1}{\rho}\nabla p + \nabla\cdot(\nu\nabla\bm{U}) - \bm{\Omega}\times\bm{U},
    \end{cases}
\label{eq_MRF_NSrotating2}
\end{equation}
which is Eq.~\eqref{eq_MRFinrotation} in the main paper. For a steady solution of Eq.~\eqref{eq_MRF_NSrotating2}, we can consider $\bm{U}$ as the unknowns that are consistent with the outer stationary region and use one unified code for the solution with some minor modifications (in terms of forcing $\bm{\Omega}\times\bm{U}$ and the $\bm{U}_R$ term leading to a correction in computing the cell face flux) for the rotating region.

\section{Data Transfer at the Interface in the RS}
\label{sec:ami:appendix}
The interfaces involved in the turbine problem are shown in Figure~\ref{fig:AMI:nonconform}. For cells on each side of the interface, the flux value on the cell face toward the interface needs to be constructed. This cannot be obtained from a standard interpolation method because the cells on two sides of the interface are non-conformal, i.e., with non-overlapping cell faces from the two sides. Given the donor and target meshes at the interface, our goal is to compute the flux value at the cell face of the target mesh. This requires reconstruction of the cells (as well as cell-centered properties) on the donor side into a ``supermesh''~\cite{farrell2009conservative} whose faces (at the interface) contain all nodes of the target mesh, so that the cell face flux on the target side can be computed by interpolation between the supermesh and target mesh. 

\begin{figure}[hbt!]
    \centering
    \includegraphics[width=0.8\textwidth]{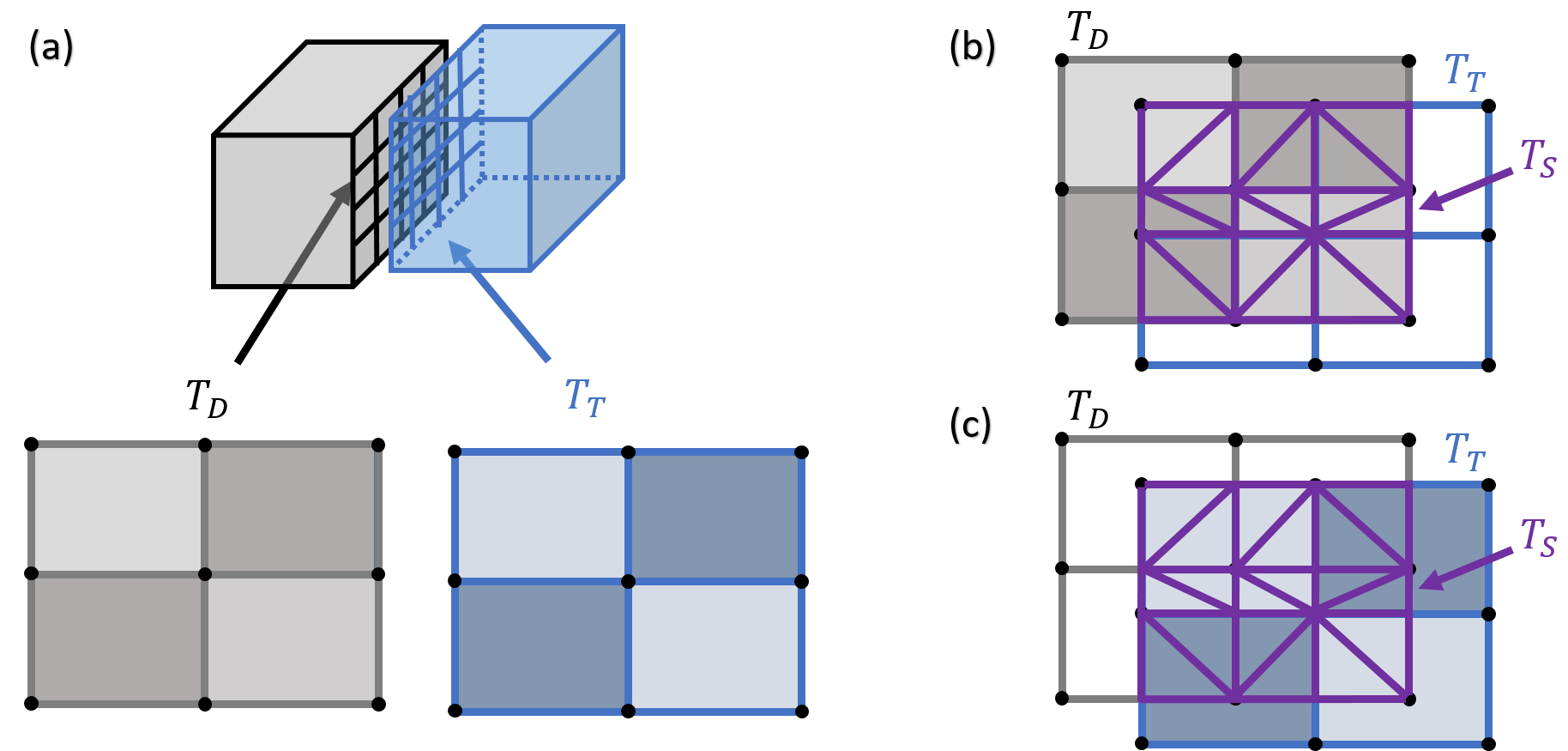}
    \caption{An example of supermesh construction at a non-conformal interface. (a) Donor mesh $T_D$ and target mesh $T_T$ are non-conformal. A triangular supermesh $T_S$ with the shaded area denoting elements (b) on $T_D$ and (c) on $T_T$.}
    \label{fig:supermesh}
\end{figure}

In general, the interpolation via supermesh involves a Galerkin projection method described in detail in~\cite{farrell2011conservative}. Here we provide a simple example to illustrate the gist of the approach, as sketched in Figure~\ref{fig:supermesh}. With donor mesh $T_D$ and target mesh $T_T$ that are non-conformal (Figure~\ref{fig:supermesh}(a)), our goal is to construct the supermesh $T_S$ that serves as a common ground for interpolation. The supermesh $T_S$ has to satisfy the properties: 
\begin{enumerate}
    \item Nodes on $T_S$ contains all the nodes on $T_D$ and $T_T$ (and intersections of their edges).
    \item For every (face cell) element in $T_S$, the intersection of it with any element of $T_D$ or $T_T$ must either be zero or the whole element.
\end{enumerate}
The connection between $T_D$ and $T_S$ is shown in Figure~\ref{fig:supermesh}(b), with the shaded area denoting cell faces of $T_D$. Since each cell face of $T_D$ is now perfectly split into triangle cell faces of $T_S$, we can assign properties to the center of the volume supermesh (associated with triangle faces) according to the area ratio of the triangles. In other words, the property at $T_D$ is distributed into fractions according to the area fraction of triangles in $T_S$. The properties at $T_S$ can then be used for interpolation with that at the target cell $T_T$ since each target cell face is also perfectly split into triangle faces of $T_S$ (Figure~\ref{fig:supermesh}(c)). Specifically, we interpolate between the volume cell of $T_S$ and the corresponding volume cell at the $T_T$ side to obtain the flux value (which is associated with the corresponding triangle area). These fluxes are then added to form the value for a single cell face of $T_T$.

\section{Analytical connection between FFD points and enclosed geometry}
\label{sec:ffd:appendix}

In this section, we explain how to obtain the analytical connection of FFD points to the enclosed geometry. While the FFD method is versatile in handling geometries in different 
dimensions, we focus on an application consistent with our work, which involves the FFD points creating a lattice box in $\mathbb{R}^3$ that encloses a three-dimensional (3D) geometry within the box. Essentially, the analytical connection is to establish a mapping $\mathbb{R}^{3} \rightarrow \mathbb{R}^3$ for each point inside the box. 
The mapping can be established by making use of the tri-variate $B$-spline function:
\begin{equation}
    X(u,v,w) = \sum_{i=0}^{N_u-1}\sum_{j=0}^{N_v-1}\sum_{k=0}^{N_w-1}{B_{i,m_u}(u)B_{j,m_v}(v)B_{k,m_w}(w)P_{i,j,k}},
\label{eq_curvesurfacevolume}
\end{equation}
where $X(u,v,w)\in \mathbb{R}^3$ is the coordinate of the volume enclosed by FFD points (and our geometry of interest is a part of this volume), parameterized by $u$, $v$, and $w$ (all $\in [0,1]$). The indexes $i$, $j$, and $k$ loop in three directions of the lattice in $\mathbb{R}^3$, $N_u$, $N_v$, and $N_w$ are the number of FFD points in each direction of the lattice, and $P_{i,j,k}\in \mathbb{R}^3$ are coordinates of the lattice FFD points. The indexes, $m_u$, $m_v$ and $m_w$ are prescribed degrees of the $B$-spline basis functions in three directions. The $B$-spline basis function is defined recursively by (taking the $i$-direction as an example)
\begin{linenomath}
\begin{gather}
    B_{i,0}(x) = \begin{cases}
                1\quad $if$ \quad t_i \leq x <t_{i+1} \\
                0\quad $otherwise$
                \end{cases},\\
    B_{i,k}(x) = \frac{x-t_i}{t_{i+k}-t_i}B_{i,k-1}(x) + \frac{t_{i+k+1}-x}{t_{i+k+1}-t_{i+1}}B_{i+1,k-1}(x),
\label{eq_Bspline}
\end{gather}
\end{linenomath}
where $t_i$ is the so-called open knot vector that is determined by choice of $N_u$ and $m_u$~\cite{kenway2010cad,reid2021free}.

With Eq.~\eqref{eq_curvesurfacevolume}, the sensitivity of the geometry to a particular FFD point, in terms of derivative, can be analytically expressed as 
\begin{equation}
    \frac{\partial X(u,v,w)}{\partial P_{i,j,k}} = B_{i,m_u}(u)B_{j,m_v}(v)B_{k,m_w}(w).
\label{eq_BsplineVolumeDerivative}
\end{equation}
Since $X\in \mathbb{R}^3$ and $P\in \mathbb{R}^3$, the derivative on the LHS of Eq.~\eqref{eq_BsplineVolumeDerivative} requires further clarification. It essentially means that the three piecewise derivatives of one $\mathbb{R}^3$ vector with respect to another $\mathbb{R}^3$ vector are equal with one another and also equal to the RHS.

In our case, linked FFD points are used as one degree of freedom (DoF). 
The derivative of the geometry coordinates with respect to the particular DoF motion can be computed by summing the contribution of each FFD point using Eq.~\eqref{eq_BsplineVolumeDerivative}.

\section{Partial Derivative Computation in the Adjoint Method}
\label{sec:adjoint:appendix}
This section presents the underlying principle of how the partial derivatives in the adjoint method can be computed. All these derivatives are computed by backward automatic differentiation~\cite{martins2021engineering} in DAFoam with graph coloring method for acceleration~\cite{gebremedhin2005color}. 

Four types of partial derivatives are involved:
\begin{equation}
    \underbrace{\frac{\partial{C_P}}{\partial\bm{s}}}_{1\times M}, \quad \underbrace{\frac{\partial{C_P}}{\partial\bm{x}}}_{1 \times N_x}, \quad \underbrace{\frac{\partial\bm{R}}{\partial\bm{s}}}_{M\times M}, \quad \underbrace{\frac{\partial\bm{R}}{\partial\bm{x}}}_{M \times N_x}.
\label{Adjoint_partialDeriv}
\end{equation}
For the former two, we consider $P=\Omega \int \bm{r} \times p d\bm{\mathcal{S}} $, where $p$ is the pressure, $\bm{r}$ is the distance to the rotating axis, the integration is over the blade surface with differential element $d\bm{\mathcal{S}}$. It is clear that $C_P$ depends on both the state variable $\bm{s}$ (i.e., pressure $p$) and design variable $\bm{x}$ (i.e., the integration surface). The derivative $\partial C_P/\partial \bm{s}$ can be simply computed by perturbing the pressure field in the flow solution. For computing $\partial C_P/\partial \bm{x}$, we write $P=\Omega \sum_{i}{\bm{r}_i\times p_i\Delta\bm{\mathcal{S}}_i}$ which is in discrete form. A perturbation in $\bm{x}$ results in perturbations of the FFD points, which deforms the geometry. This, in turn, leads to the surface mesh deformation that affects the $\bm{r}_i$ and $\Delta\mathcal{S}_i$. The partial derivative $\partial C_P/\partial \bm{x}$ can therefore be computed using the chain rule to connect all processes.  

For the latter two, we consider the discretized governing equation in residual form, i.e. $\bm{R}=0$. Perturbation of $\bm{s}$ changes the velocity field, pressure field, and the velocity flux (at cell faces) so that $\bm{R}$ is perturbed. The derivative $\partial\bm{R}/\partial\bm{s}$ can then be computed correspondingly. Perturbation of $\bm{x}$ leads to the deformation of all computational mesh in the full fluid domain (discussed in Section~\ref{subsubsec:meshdeform}). The construction of velocity flux (or more precisely, the coefficients in the algebraic governing equation) is thus affected, leading to the computation of $\partial\bm{R}/\partial\bm{x}$ correspondingly.

\section{Discrete grid-search in a lower-dimensional design space}
\label{sec:discreteDesign:appendix}

\begin{figure}[hbt!]
    \centering
    \includegraphics[width=0.8\textwidth]{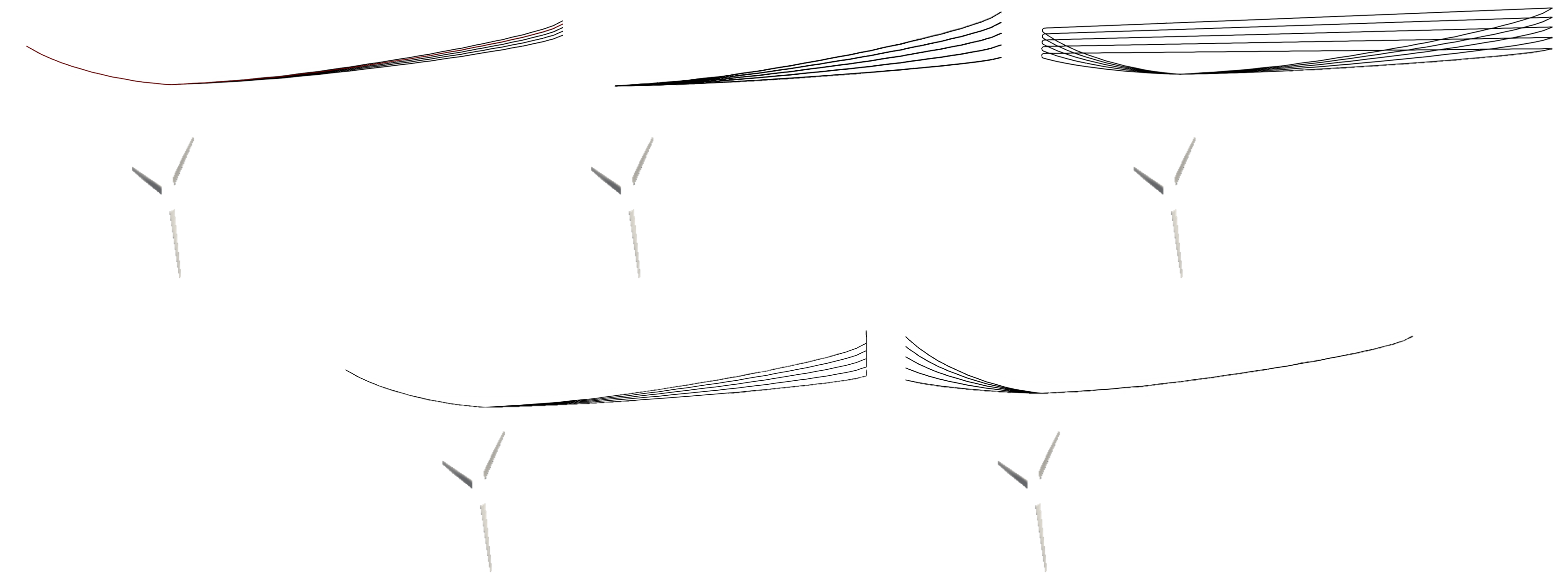}
    \caption{Five sets of duct designs in the grid-search study.}
    \label{fig:bruteforce_design}
\end{figure}

This section describes our effort in a brute-force grid search of a lower-dimensional design space. Taking the baseline duct design as a reference~\cite{Knight2018}, we first create 5 sets of duct configurations, each set representing one type of duct with 5 varying designs each. As sketched in Figure~\ref{fig:bruteforce_design}, the 5 sets are: (1) varying outlet area from the baseline design; (2) varying outlet area from the baseline design with throat section as the duct inlet; (3) varying both inlet and outlet areas from the baseline design with a finite-thickness duct (outer surface simply as a straight line); (4) varying outlet area from the baseline design with a flange (inspired by \citet{ohya2010shrouded}); (5) varying inlet area from the baseline design. In total, these represent 25 duct designs, with the ratio of maximum and minimum duct areas ranging from 1.25 to 2.75. For each of these duct designs, other design variables considered are listed below with discrete values:
\begin{itemize}
    \item $\lambda$: $3,\:3.5,\:4,\:4.5,\:5,\:5.5,\:6,\:6.5,\:7,\:7.5$
    \item Blade root pitch: $10^{\circ},\: 20^{\circ},\: 30^{\circ},\: 40^{\circ},\: 45^{\circ}$
    \item Twist profiles: Baseline blade designs A and B
    \item Tip gap ratio: $0.09R,\: 0.2R$
\end{itemize} 

\begin{figure}[hbt!]
    \centering
    \includegraphics[width=0.7\textwidth]{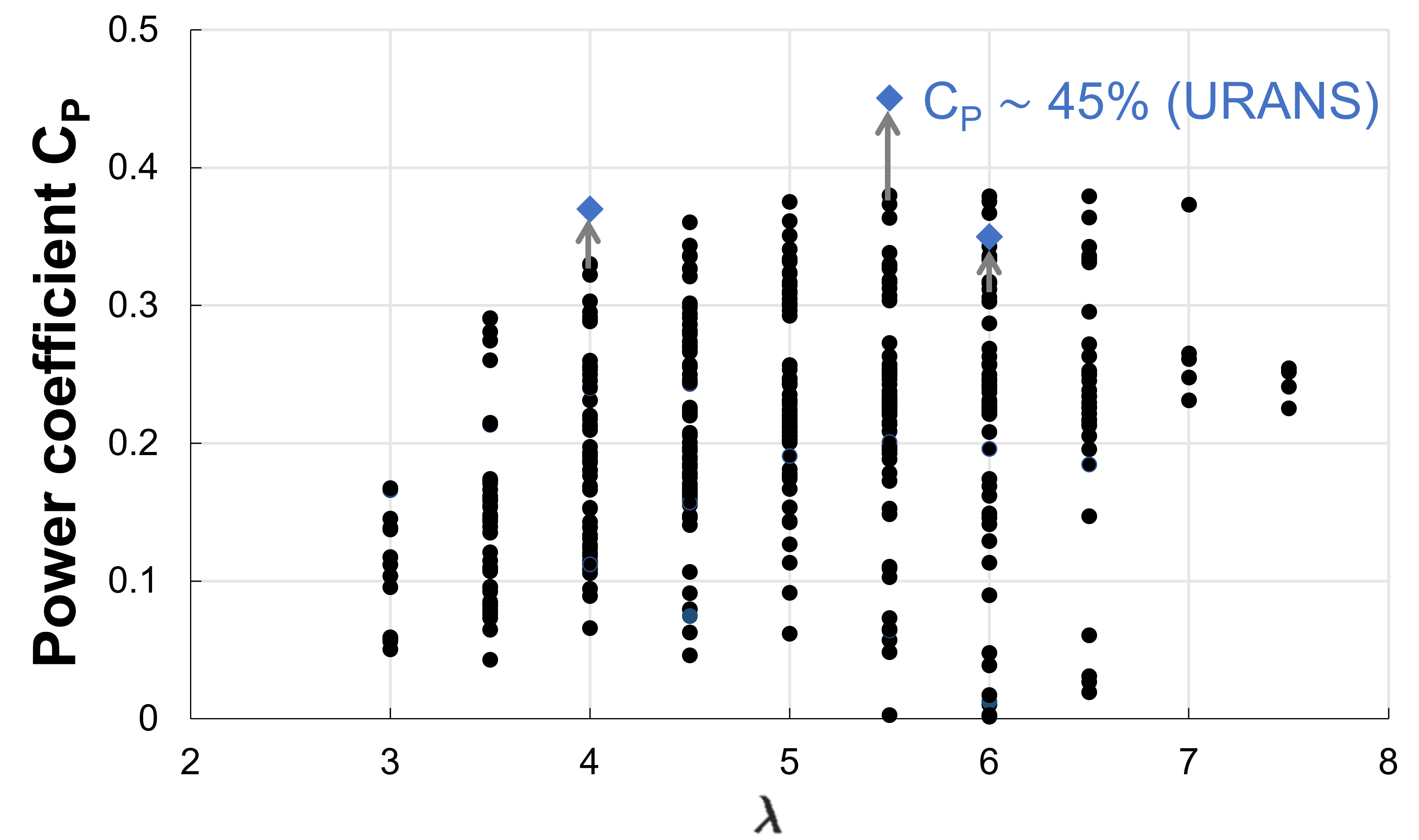}
    \caption{$C_P$'s obtained from $\mathcal{O}(450)$ the RANS-MRF simulations for discrete points in the design space, with results stacked to each $\lambda$. The URANS re-evaluations of some designs with relatively high $C_P$'s from the RANS-MRF yield results shown by diamonds.}
    \label{fig:bruteforce}
\end{figure}

The RANS-MRF solver is used for each design to evaluate its performance on a coarse grid with 1-2 million cells (even coarser than M0). In total, about 450 cases are run with results of their efficiency $C_P$ shown in Figure~\ref{fig:bruteforce}.
Since it is difficult to visualize results in the design space of more than three dimensions, we simply plot all results as functions of $\lambda$, i.e., for each $\lambda$, a large number of $C_p$ values resulting from varying other design variables are stacked. While it is not our goal to further distinguish designs at each $\lambda$, it is clear that the maximum $C_p$ among all $\mathcal{O}(450)$ cases is only 38\% evaluated by the RANS-MRF. We pick up a few designs with relatively high $C_p$ and re-evaluate their performances using the higher-fidelity URANS solver on the M0 grid, with results also shown in Figure~\ref{fig:bruteforce}. We see that the maximum efficiency computed by the URANS is 45\%, which happens to correspond to our baseline design B (in terms of both the duct and blade geometry) but is still lower than 46\% from the unducted Bahaj model. Therefore, a simple brute-force grid search, as performed here, does not provide any duct turbine design with higher efficiency than that of standard unducted turbines.

 \bibliographystyle{elsarticle-num-names} 
 \bibliography{cas-refs}

\end{document}